\documentclass[11pt,reqno]{amsproc}
\usepackage[margin=1in]{geometry}
\usepackage{amsmath, amsthm, amssymb}
\usepackage{times, esint,stackrel}
\usepackage{enumitem}
\usepackage{color}
\usepackage{hyperref}
\usepackage{mathabx}
\usepackage{mathrsfs}

\newcommand{\be}{\begin{equation}}
\newcommand{\ee}{\end{equation}}
\newcommand{\bea}{\begin{eqnarray}}
\newcommand{\eea}{\end{eqnarray}}
\newcommand{\lb}{\label}

\newtheorem{theorem}{Theorem}

\newtheorem{lemma}{Lemma}

\newtheorem{rem}{Remark}

\newcommand{\bx}{{\mathbf{x}}}

\newcommand{\bk}{{ \mathbf{k}}}

\newcommand{\bv}{{ \mathbf{v}}}

\newcommand{\bu}{{ \mathbf{u}}}

\newcommand{\mpr}{P}
\newcommand{\bsf}{{\mathbf{f}}}

\newcommand{\bph}{{\boldsymbol{\varphi}}}
\newcommand{\bvp}{{\mathbf{v}}}
\newcommand{\bz}{{ \mathbf{z}}}
\newcommand{\bxx}{{\boldsymbol{\xi}}}

\newcommand{\grad}{{\mbox{\boldmath $\nabla$}}}
\newcommand{\bdot}{{\mbox{\boldmath $\cdot$}}}

\newcommand{\bzed}{{\mbox{\boldmath $0$}}}

\newcommand{\botau}{{\mbox{\boldmath $\tau$}}}
\newcommand{\boeta}{{\mbox{\boldmath $\eta$}}}

\title[Statistical Euler Solutions]{Space-Time Statistical Solutions 
of the Incompressible Euler Equations and Landau-Lifschitz Fluctuating Hydrodynamics}
\author{Gregory L. Eyink and Lowen Peng}
\address{Department of Applied Mathematics and Statistics,  Johns Hopkins University, Baltimore, MD 21218}
\email{eyink@jhu.edu, lpeng22@jhu.edu}

\date{today}
\begin{document}

\begin{abstract}

We study the infinite Reynolds limit of the solutions of the Landau-Lifschitz-Navier-Stokes equations for an incompressible fluid on a $d$-dimensional torus for $d\geq 2$. These equations, which model thermal fluctuations in fluids, are given a standard physical interpretation as a low-wavenumber ``effective field theory'', rather than as stochastic partial differential equations. We study solutions which enjoy some Besov regularity in space, as expected for initial data chosen from a driven, turbulent steady-state ensemble. The empirical basis for this regularity hypothesis, uniform in Reynolds number, is carefully discussed. Considering the initial-value problem for the Landau-Lifschitz equations, our main result is that the infinite Reynolds number limit is a space-time statistical solution of the incompressible Euler equations in the sense of Vishik \& Fursikov. Such solutions are described by a probability measure on space-time velocity fields whose realizations are weak solutions of the incompressible Euler equations, each with the same prescribed initial data.\\ 


\end{abstract}

\maketitle

\section{Introduction}

The main motivation for the present work arises in the theory of {\it spontaneous stochasticity} 
for turbulent flows \cite{falkovich2001particles}. Although the pre-history of this notion lies in the 
classic study of Richardson on 2-particle turbulent dispersion \cite{richardson1926atmospheric}, 
the concept was first clearly formulated in the work of Bernard, Gaw\c{e}dzki, and Kupiainen 
\cite{bernard1998slow}. Their fundamental observation was that the H\"older exponent 
of the infinite-Reynolds velocity field must be $\leq 1/3,$ according to a theorem of Onsager 
\cite{onsager1949statistical}, and this regularity is too low to guarantee uniqueness of the 
solutions to the ODE for Lagrangian particle trajectories. The authors of \cite{bernard1998slow} showed 
that this non-uniqueness manifests in remnant stochasticity of Brownian particles (e.g. dye molecules)
in the limit of large Reynolds and P\'eclet numbers, at least for the soluble model of Kraichnan 
\cite{kraichnan1968small}, even though the limiting evolution equations for a fixed velocity realization are deterministic. 
Furthermore,  they showed that such ``Lagrangian spontaneous stochasticity'' is necessary in general for a 
scalar advected by any turbulent flow, either passively or actively, to exhibit anomalous dissipation
in the infinite Reynolds \& P\'eclet limit \cite{drivas2017lagrangian}. More recently, it was pointed out by 
Mailybaev \cite{mailybaev2016spontaneously,mailybaev2016spontaneous} that there is a closely 
allied phenomenon of ``Eulerian spontaneous stochasticity'' in which the entire velocity field 
obtained by solving the deterministic Navier-Stokes equations remains random in the infinite-Reynolds 
limit, even for initial data with vanishing randomness. This result was anticipated in the profound work 
of Lorenz \cite{lorenz1969predictability} on turbulence predictability, who argued that ``certain formally 
deterministic fluid systems which possess many scales of motion are observationally indistinguishable 
from indeterministic systems.''  Spontaneous stochasticity provides a precise mathematical formulation 
of Lorenz' ideas, in which the ``formally deterministic fluid system'' exhibiting random solutions
is governed by the ideal Euler equations with fixed initial data, obtained in the simultaneous limit of infinite 
Reynolds number and vanishing noise. The mathematical fact which underlies this persistent randomness 
is the non-uniqueness of solutions to the Cauchy initial-value problem for the limiting Euler equations, 
even with additional admissibility constraints imposed such as H\"older regularity and local dissipativity 
\cite{de2010admissibility,de2017high,daneri2021non}. The prototypical example of a Kelvin-Helmholtz
unstable vortex sheet provides strong empirical evidence for this phenomenon 
\cite{szekelyhidi2011weak,thalabard2020butterfly}. 
 
The vanishing noise considered in the work of Lorenz was small-scale external perturbations, such as the proverbial flap of a butterfly's wing
\cite{palmer2014real}. Intrinsic thermal noise has been more  recently investigated, following 
the pioneering work of Betchov \cite{betchov1957fine,betchov1961thermal} and Ruelle 
\cite{ruelle1979microscopic}. The distinct feature of thermal noise is that it is unavoidable, since 
it is a necessary consequence of the local thermodynamic equilibrium required for validity of a 
hydrodynamic description. The simplest fluid dynamical model which incorporates 
thermal fluctuation effects is Landau-Lifschitz fluctuating hydrodynamics \cite{landau1959fluid},
a stochastic modification of the Navier-Stokes equations. For incompressible fluids in the limit 
of low Mach numbers, this model takes the form in any space dimension $d\geq 2$ 
\be \partial_t\bu + \pi_\Lambda(\bu\cdot\grad)\bu = -\grad p + \nu_\Lambda\triangle \bu 
+ \sqrt{\frac{2\nu_\Lambda k_B T}{\rho}} \grad\cdot \boeta, \qquad \grad\bdot\bu =0, 
\lb{FNS2} \ee
where $\pi_\Lambda$ is the projection to Fourier modes with wavenumbers $|\bk|<\Lambda,$ 
$\nu_\Lambda$ is (scale-dependent) kinematic viscosity, $T$ is absolute temperature, $k_B$ is 
Boltzmann's constant, $\rho$ is mass density, and $\botau=(2\nu_\Lambda k_B T/\rho)^{1/2} 
\boeta$ is a fluctuating stress modeled as a Gaussian 
random matrix field, both symmetric and traceless, with mean zero and covariance
\begin{eqnarray}
\langle \eta_{ij}(\bx,t) \eta_{kl}(\bx',t')\rangle& = &
\left(\delta_{ik}\delta_{jl}+\delta_{il}\delta_{jk} -\frac{2}{3}\delta_{ij}\delta_{kl}\right)
\delta^d_\Lambda(\bx-\bx')\delta(t-t'). \lb{FDR}
\end{eqnarray}
The kinematic pressure $p$ is obtained from the incompressibility constraint $\grad\bdot\bu =0.$
For example, see \cite{forster1977large,donev2014low}. Although the model in \eqref{FNS2},\eqref{FDR}  
appears superficially to be given 
by a stochastic partial differential equation, such a continuum interpretation is not the customary one in 
the physics literature. Instead, the model is understood to contain an explicit high-wavenumber cutoff 
$\Lambda,$ which may be freely chosen in the range $\ell_\nabla^{-1}\ll \Lambda\ll \lambda_{mfp}^{-1}$ 
where $\ell_\nabla$ is the gradient length of the flow and $\lambda_{mfp}$ is the molecular mean free path 
length of the fluid. In the 
language of theoretical physics, the model is an ``effective field theory'' which describes molecular fluids only at 
large length-scales $>|\Lambda|^{-1}$ \cite{schwenk2012renormalization,liu2018lectures}. 
In particular, the subscript on $\delta^3_\Lambda$ in \eqref{FDR} 
indicates that this is not a true delta-function, but only an effective delta-function for band-limited 
functions restricted to wavenumbers $|\bk|<\Lambda.$ For more discussion, see Section \ref{sec:back}. 

Simulations of turbulence in the presence of thermal noise have detected sensitivity 
of large-scale flow features in individual flow realizations to such noise, even at high Reynolds
numbers \cite{gallis2021turbulence}. The prior work of one of us \cite{bandak2023spontaneous}
studied systematically the limit of 
high Reynolds number $Re=UL/\nu_\Lambda$ for the dynamics \eqref{FNS2}, which, after non-dimensionalization 
based on large-scale flow parameters, takes the form 
\begin{equation}
\partial_t\bu + \pi_{\hat{\Lambda}}(\bu\cdot\grad)\bu = -\grad p + \frac{1}{Re}\triangle \bu 
+ \sqrt{\frac{2\theta_\eta}{Re^{3(d+2)/4}}} \grad\cdot \boeta + \digamma {\mathbf f}. \lb{FNS3} 
\end{equation}
See section \ref{sec:back} for detailed discussion. Here we note only that $\theta_\eta$ 
is the ratio of thermal energy to the energy of a Kolmogorov-scale eddy, which is a 
generally tiny quantity of order $10^{-8}$ or smaller and fixed as $Re\to\infty.$ 
We have added as well a large-scale body force ${\bf f}$ to drive turbulence, with a dimensionless
amplitude $\digamma$ which is also fixed as $Re\to\infty.$ It is important to note, however, 
that the dimensionless wavenumber cutoff $\hat{\Lambda}=L\Lambda$ diverges together with $Re,$
e.g. $\hat{\Lambda}\,\propto \,Re^{3/4}$ if the cutoff is fixed relative to the Kolmgorov dissipation 
scale $\eta$ or $\hat{\Lambda}\propto Re$ if instead the cutoff is fixed relative to the 
molecular mean free path $\lambda_{mfp}.$ Therefore, formally, the limiting dynamics 
as $Re\to\infty $ is described by the forced continuum Euler equations 
\begin{equation}
\partial_t\bu + \grad\bdot(\bu\otimes \bu) = -\grad p + \digamma {\mathbf f}, \lb{Euler} 
\end{equation}
with both the viscous diffusion and the thermal fluctuations vanishing in the limit. To investigate 
numerically the limiting behavior for $Re\gg 1,$ the authors of \cite{bandak2023spontaneous} 
studied a simplified ``shell model'' version of eq.\eqref{FNS2} which allowed them to simulate 
a broad range of Reynolds numbers $10^6$--$10^8.$ For quasi-singular initial data ${\bf u}_0$ with 
K41 spectra over a broad range of wave-numbers, or with even more singular initial data ${\bf u}_0$
selected from a steady-state turbulent simulation, the transition probability $P_{Re}({\bf u}, t|{\bf u}_0, t_0) $ 
was found to become $Re$-independent for sufficiently large Reynolds number. Most remarkably, the limiting 
statistics were found to be very universal, independent of the selected sequence $Re_k\to\infty,$
the precise details of the thermal noise model, etc. This universality is one of the hallmarks
of spontaneous stochasticity \cite{thalabard2020butterfly,eyink2020renormalization,mailybaev2022spontaneous}. 

The empirical observation by \cite{bandak2023spontaneous} of non-trivial infinite-$Re$ limits  
of transition probabilities for fixed initial data ${\bf u}_0,$ strongly suggests the existence 
of non-unique solutions of the Cauchy problem for the limiting Euler equation \eqref{Euler}. 
Such a non-uniqueness is consistent with the recent results from convex integration constructions, 
\cite{de2010admissibility,de2017high,daneri2021non}, that provide a mathematical underpinning 
for spontaneous stochasticity. However, the Euler solutions provided by convex integration 
do not correspond to infinite-Reynolds limits. Furthermore, no mathematical justification 
was offered in the numerical study of high-$Re$ turbulence 
\cite{bandak2023spontaneous} that the limiting 
velocity realizations should indeed be (weak) solutions of the ideal Euler equation \eqref{Euler}. 
The purpose of the present paper is to fill this gap and to prove that infinite-$Re$
limiting probability measures exist for the Landau-Lifschitz dynamics \eqref{FNS3} that are 
supported indeed on such weak solutions of the Euler equations. We shall model our proofs closely 
after the work of Vishik \& Fursikov \cite{vishik1988mathematical}, who generalized Leray's 
construction of weak solutions of the incompressible Navier-Stokes equation to space-time
statistical solutions of incompressible Navier-Stokes equation with random initial data. 
In fact, the monograph \cite{vishik1988mathematical} developed a very general approach, which can be applied to a 
great many models described by PDE's or stochastic PDE's. 
Unlike the work on Navier-Stokes, however, we cannot derive the necessary regularity for 
the limit from {\it a priori} energy estimates. As is well-known, the compactness that 
follows from kinetic energy bounds alone does not suffice to yield standard weak Euler 
solutions in the infinite-$Re$ limit, but instead only still more generalized notions 
of Euler solution such as the ``dissipative solutions'' in the sense of Lions 
\cite{lions1996mathematical} or measure-valued weak solutions \cite{diperna1987oscillations}. 
While it cannot be ruled out that such generalized weak solutions are physically relevant
\cite{fjordholm2016computation,thalabard2020turbulence}, it was pointed out by Chen \& Glimm 
\cite{chen2012kolmogorov} that standard weak solutions can be obtained in the infinite-$Re$ limit 
if additional conditions are satisfied that may be empirically verified. See also following work  
\cite{constantin2018remarks,drivas2019onsager,drivas2019remarks,isett2022nonuniqueness}.
We follow a version of this strategy here, invoking a condition of spatial Besov regularity uniform 
in the Reynolds number that was first employed to study inviscid limits in \cite{drivas2019onsager} 
but which in fact goes back to a remark of Isett \cite{isett2022nonuniqueness}. 
Recently, Fjordholm et al. \cite{fjordholm2024vanishing} have proved an inviscid limit result with a similar assumption, but in the framework of statistical solutions in the Foia\c{s}-Prodi sense, which involves single-time measures on spatial velocity fields. 
See Remark \ref{rem1} below. 

{\it Theorem Statement}. We now formulate in precise mathematical terms the main result of this work. 
We take as flow domain the $d$-dimensional torus ${\mathbb T}^d=[0,2\pi)^d,$ $d\geq 2,$ in order to avoid issues
with boundaries. Indeed, the proper formulation of fluctuating hydrodynamics in the presence of solid walls 
is still a subject under active investigation \cite{reichelsdorfer2016foundations}. We consider the incompressible 
fluid velocity field $\bu^\nu$ governed by the {\it Landau-Lifshitz-Navier-Stokes} (LLNS) equations written
in the dimensionless form \eqref{FNS3}, or 
\begin{equation}\label{llns}
\mathrm{d}\bu^\nu=\mathscr{P}\left[-\grad\cdot \pi_\Lambda(\bu^\nu\otimes\bu^\nu)+\nu\triangle\bu^\nu+\pi_\Lambda\bsf\right]\mathrm{d}t+\nu^{\kappa}\mathscr{P}\grad\cdot \pi_\Lambda
\mathrm{d}{\bxx}, \quad \theta>0\end{equation}
where we follow standard practice in the mathematics literature of denoting $\nu=1/Re$ and considering
the infinite Reynolds number limit as an ``inviscid limit.'' We have, for simplicity, set $\theta_\eta=\digamma=1,$
since these finite constants are fixed in the limit $\nu\to 0$ and play no essential role in the argument. The 
projection $\pi_\Lambda$ is defined in terms of the Fourier transform $(\mathcal{F}f)(\bk)=\frac{1}{(2\pi)^d}\int_{{\mathbb T}^d} e^{-i\bk\bdot\bx} f(\bx) \, \mathrm{d}^dx$ by 
\be \pi_\Lambda f(\bx)=\sum_{\bk\in{\mathbb Z}^d,\, |\bk|_\infty<\Lambda} e^{i\bk\bdot\bx} (\mathcal{F}f)(\bk),  \lb{proj} \ee 
where for technical reasons of the proof related to Lizorkin representation of function spaces, we 
use the max-norm $|\bk|_\infty=\max_{i=1}^d |k_i|$ and $\Lambda=2^m$ with $m(\nu)\in {\mathbb N}$
such that $\Lambda\sim \nu^{-\alpha}$ as $\nu\to 0,$ $3/4\leq \alpha\leq 1.$ The proof would work for a wider range
of $\alpha$ but, as discussed in the Introduction and in Section \ref{sec:back}, only the above range is of 
physical relevance. We have eliminated the pressure which enforces incompressibility in \eqref{FNS3} by the Leray--Hodge projection 
$\mathscr{P}\equiv\operatorname{I}-\grad\triangle^{-1}\operatorname{div},$ given in Fourier representation by 
\[(\mathcal{F}\mathscr{P}\bu)_i(\mathbf{k})=\sum_{j=1}^d\left(\delta_{ij}-\frac{k_i k_j}{|\mathbf{k}|^2}\right)(\mathcal{F}\bu)_j(\mathbf{k}),
\quad i=1,\ldots,d. \]
To represent the random forcing by thermal noise, we take a probability space $(\Omega,\mathscr{F},{\mathbb P}),$ an increasing 
filtration $(\mathscr{F}_t: \, t\in [0,T])$ and an $\mathscr{F}_t$-adapted cylindrical Brownian motion $\bxx$ on the Hilbert space $L^2({\mathbb T}^d)$ \cite{brehier2014short}, 
with $\langle \bxx(t),h\rangle$ for $t\in [0,T],$ $h\in L^2({\mathbb T}^d)$ taking values in the space of symmetric, traceless $d\times d$ matrices,
with mean ${\mathbb E}[\langle \bxx(t),h\rangle]=\bzed$ and covariance 
$$ {\mathbb E}[\langle \xi_{ij}(t),h\rangle\langle \xi_{kl}(t'),h'\rangle]=
\left(\delta_{ik}\delta_{jl}+\delta_{il}\delta_{jk} -\frac{2}{3}\delta_{ij}\delta_{kl}\right)\langle h,h'\rangle(t\wedge t'). $$
The cylindrical Brownian motion can be represented \cite{brehier2014short} as $\langle \bxx(t),h\rangle=\frac{1}{(2\pi)^d}\int_{{\mathbb T}^d}\bxx(\bx,t) h(\bx)\,d^dx$
with 
\be \bxx(\bx,t) = \sum_{\bk\in {\mathbb Z}^d} \xi_{ij}^\bk(t) e^{i\bk\bdot\bx} \lb{xidef} \ee
where $\xi_{ij}^\bk(t)$ for $\bk\in {\mathbb Z}^d$ are an infinite set of complex Brownian motions on the probability space $(\Omega,\mathscr{F},{\mathbb P}),$
with covariance 
$$ {\mathbb E}[\xi_{ij}^\bk(t)\xi_{kl}^{\bk'}(t')]=
\left(\delta_{ik}\delta_{jl}+\delta_{il}\delta_{jk} -\frac{2}{3}\delta_{ij}\delta_{kl}\right) 
\delta_{\bk+\bk',\bzed}(t\wedge t'), $$
and thus independent except for the constraints $\overline{\xi_{ij}^\bk(t)}=\xi_{ij}^{-\bk}(t),$ $\xi_{ji}^\bk(t)=\xi_{ij}^\bk(t)$ and 
$\sum_{i=1}^d \xi_{ii}^\bk(t)=0.$ The representation \eqref{xidef} is particularly useful in conjunction with the projection $\pi_\Lambda$
which simply restricts the summation over $\bk$ to $|\bk|_\infty \leq \Lambda.$ The prefactor $\nu^{\kappa}$ of the stochastic term 
originates from careful non-dimensionalization of \eqref{FNS2} with Taylor's relation (see Section \ref{sec:back}), but our proof 
would work with any factor vanishing as $\nu\to 0$. We include also in \eqref{llns} a deterministic stirring force 
$\bsf$ to agitate turbulence. In addition to some mild regularity which allows for existence of solutions to \eqref{llns}, we require also 
$$ ({\mathcal F}\bsf)(\bzed)= \frac{1}{(2\pi)^d}\int_{{\mathbb T}^d} \bsf(\bx) \, \mathrm{d}^dx=\bzed, $$
so that there is no bulk acceleration of the fluid. 

The physical necessity of the high-wavenumber cutoff $\Lambda$ is crucial for the solution theory of the Landau-Lifschitz
equations. In fact, the equation \eqref{llns} is not an SPDE but instead a finite set of It$\bar{\rm o}$ SDE's with an energy-conserving
quadratic nonlinearity, of a type very well-studied mathematically. For example, the lecture notes of Flandoli \cite{flandoli2008introduction}, Section 3, 
show that strong stochastic solutions $\bu^\nu$ of \eqref{llns} exist globally in time for $\bsf=\bzed$ and are pathwise unique, 
at any $\nu>0.$ In fact, under very modest regularity assumptions, the same is true with a force $\bsf\neq \bzed.$ Thus,
for fixed initial data $\bu_0^\nu=\pi_\Lambda\bu_0$, there is a well-defined distribution $P^\nu_{\bu_0}$ on solutions of \eqref{llns} induced 
by the measure ${\mathbb P}.$ If the initial data satisfies the zero-momentum condition 
\be ({\mathcal F}\bu_0)(\bzed)= \frac{1}{(2\pi)^d}\int_{{\mathbb T}^d} \bu_0(\bx) \, d^dx=\bzed, \lb{meanu0} \ee
then it is furthermore easy to show \cite{flandoli2008introduction} that 
\be ({\mathcal F}\bu)(\bzed,t)= \frac{1}{(2\pi)^d}\int_{{\mathbb T}^d} \bu(\bx,t) \, d^dx=\bzed, \quad \forall t\in [0,T],\quad P^\nu_{\bu_0}-a.e. \ \bu \lb{meanu} \ee 
For the purposes of studying spontaneous stochasticity, we are interested in non-smooth initial velocity data $\bu_0\notin C^{1,\alpha}({\mathbb T}^d),$ 
$0<\alpha<1,$ for which local existence of a smooth Euler solution is not guaranteed. This restriction would not be necessary, if the smooth Euler 
solution should develop a singularity in finite-time for initial data $\bu_0\in C^{1,\alpha}(\mathbb{T}^d).$ However, singular initial data are natural,
for example, if the velocity field $\bu_0$ is chosen from a steady-state turbulent ensemble at $Re=\infty.$ More modest regularity such as 
$\bu_0\in L^p({\mathbb T}^d),$ $p>2$ is nevertheless expected for such initial data, which can in fact be shown \cite{flandoli2008introduction}
to be preserved by the dynamics \eqref{llns}. We are thus motivated more generally to consider initial data $\bu_0$ distributed 
according to a Borel measure $\mu$ on $L^p({\mathbb T}^d)$ and satisfying \eqref{meanu0} $\mu$-a.s. We denote as $P^\nu_{\mu}$ 
the distribution over the solutions of \eqref{llns} induced by the product probability measure ${\mathbb P}\times \mu$ and, for simplicity, 
we denote expectation over the latter measure by ${\mathbb E}_\mu.$ The initial data of primary interest for the study 
of spontaneous stochasticity are Dirac measures of the form $\mu=\delta_{\bu_0},$ but we may always restrict attention subsequently 
to such a special case.  

A principal ingredient of our theorem is the following {\it hypothesis} on the family of distributions $P^\nu_\mu$: 
\be \sup_{\nu>0}\int\lVert\bu\rVert_{L^r([0,T],B_{p,\infty}^{\sigma}(\mathbb{T}^d))}^2\,\mathrm{d}\mpr^\nu_\mu(\bu)<\infty,  \lb{besov} \ee 
with $2<r,p<\infty,$ $0<\sigma<1.$  This condition is a probabilistic version of the assumption of uniform boundedness of Besov norms 
for incompressible Navier-Stokes solutions $\bu^\nu\in L^3([0,T],B_{3,\infty}^{\sigma}(\mathbb{T}^d)),$ invoked in several works 
\cite{drivas2019onsager,drivas2019remarks,isett2022nonuniqueness} to derive sufficient compactness of the sequence $\bu^\nu$ 
to guarantee strong convergence of suitable subsequences to weak Euler solutions. We are far from deriving such regularity 
{\it a priori} (but see \cite{drivas2022self}) and, as in those previous works, the primary justification is experimental and numerical observations 
on turbulent velocity fields. We shall discuss in the following section \ref{sec:back} the available empirical evidence that uniform Besov regularity \eqref{besov} holds even in the presence of thermal noise. 

With this background, we can now state our main result: 

\begin{theorem}\label{result2} Let $\mpr^\nu_\mu,$ $\nu>0$ be the distribution on zero-space-mean It$\bar{o}$ strong solutions to (\ref{llns}) for $d\geq 2$, supposing 
that the deterministic, zero-mean forcing has regularity $\bsf\in C^0([0,T]\times \mathbb{T}^d)$ and that the initial distribution $\mu$ 
is supported on $L^p(\mathbb{T}^d)$, $p>2$. We assume that the initial data satisfy $\grad\,\bdot\,\bu_0=0$ in the distributional sense, for
$\mu-$a.e. $\bu_0.$ We furthermore assume the uniform regularity condition \eqref{besov} for 
the same $p$ and $2<r<\infty,$ $0<\sigma<1,$ with $\Lambda=2^{m(\nu)}\sim \nu^{-\alpha}$ for $m:(0,\infty)\to\mathbb{N}$ and $3/4\leq\alpha\leq1$. 
Then the following statements hold: \\
(i) The family of measures $\mpr^\nu_\mu,$ $\nu>0$ is tight on the Borel $\sigma$-algebra ${\mathcal B}(\mathcal{Z}^{\sigma,\beta}_{r,p})$
of the Banach space  
\be \mathcal{Z}^{\sigma,\beta}_{r,p}:=L^r([0,T],L^p(\mathbb{T}^d))\cap C^0([0,T],B_{p/2,\infty}^{\sigma-\beta}(\mathbb{T}^d)), \quad
\beta\geq1+\sigma+2d. 
\lb{Zdef} \ee
Thus, there exists a subsequence $\nu_k\to 0$ such that $\mpr^{\nu_k}_\mu$ converges weakly to some measure $\mpr_\mu,$ i.e.
\[\int\varphi\,\mathrm{d}\mpr^{\nu_k}_\mu\to\int\varphi\,\mathrm{d}\mpr_\mu \ \mbox{ as } \ \nu_k\to 0, \quad \forall\varphi\in C^0_b(\mathcal{Z}^{\sigma,\beta}_{r,p}), \] 
the latter space consisting of functions $\varphi$ bounded and continuous in the strong norm topology on $\mathcal{Z}^{\sigma,\beta}_{r,p}.$ \\
(ii) $P_\mu$ is supported on 
$$\widetilde{\mathcal{E}}^{\sigma,\beta}_{r,p}:=
L^r([0,T],B_{p,\infty}^{\sigma}(\mathbb{T}^d))\cap C^0([0,T],B_{p/2,\infty}^{\sigma-\beta}(\mathbb{T}^d))$$
and furthermore the zero-mean condition holds, so that 
\[ \int_{{\mathbb T}^d} \bu(\bx,t) \, \mathrm{d}^dx=\bzed, \quad \forall  t\in[0,T] \qquad  P_\mu-a.e.\ \bu.  \lb{meanu-P} \] 
\\
(iii) $P_\mu$ is supported on standard weak solutions of the incompressible Euler equations, so that distributionally in space-time 
\[   \partial_t \bu =\mathscr{P}[-\grad\cdot(\bu\otimes \bu)+\bsf],\quad  \grad\bdot\bu=0, \qquad P_\mu-a.e.\ \bu \]\\
(iv) Because of (ii), the trace $\gamma_{0}:\bu\mapsto\bu(\cdot,0)$ to initial data is well-defined as a continuous map
\[\gamma_0:\mathcal{Z}^{\sigma,\beta}_{r,p}\to B_{p/2,\infty}^{\sigma-\beta}(\mathbb{T}^d)\]
and the pushfoward measure $\mpr^{(0)}_\mu=\gamma_{0\ast}\mpr_\mu$ is defined on $\mathcal{B}(B_{p/2,\infty}^{\sigma-\beta}(\mathbb{T}^d))$.
In fact, $\mpr^{(0)}_\mu$ is supported on $L^p({\mathbb T}^d)$ and when restricted to $\mathcal{B}(L^p(\mathbb{T}^d))$ it coincides with $\mu.$
\end{theorem}

We make a sequence of remarks concerning this result: 

\begin{rem}\lb{rem1} 
{\rm Important physical consequences of the theorem are obtained for the special case of deterministic (Dirac delta) 
initial data, $\mu=\delta_{\bu_0}.$ In that case, the implication is that any measure $P_{\bu_0}$ obtained 
as a weak limit for $\nu\to 0$ is supported on weak Euler solutions $\bu,$ all of which have the same initial 
datum $\bu(\cdot,0)=\bu_0.$ If the measure $P_{\bu_0}$ is non-trivial for suitable non-smooth initial data $\bu_0,$ 
as suggested by numerical evidence \cite{bandak2023spontaneous}, then such non-triviality requires non-unique
solutions of the corresponding Cauchy problem for Euler. This is the essence of the concept of ``spontaneous stochasticity''. Our proof can be easily modified for the stochastic shell models employed in the numerical studies
\cite{bandak2023spontaneous} and, in fact, the additional regularity assumption \eqref{besov} should be unnecessary in that case \cite{constantin2007regularity}. We remark that the same implications
cannot be drawn using the method of statistical solutions of Foia\c{s}-Prodi, as in the recent 
work \cite{fjordholm2024vanishing}. The statistical Euler solutions constructed in that work
are probability measures $P_\mu^{(t)}$ on a function space such as $L^2({\mathbb T}^d)$ which are parameterized by time $t\in[0,T]$ and which satisfy, in a weak suitable sense, the forward Kolmogorov 
or Liouville functional equation for an ensemble of incompressible Euler equations. Such statistical 
solutions should be obtained from our space-time statistical solutions as pushforward measures
$P_\mu^{(t)}=\gamma_{t*}P_\mu$ where $\gamma_t:\bu\mapsto \bu(\cdot,t)$ is the time-$t$ trace.
For rigorous results of such type, see \cite{vishik1988mathematical}, Theorem XI.1.1. It is likely 
that the approach of \cite{fjordholm2024vanishing} can be extended to the inviscid limit of the Landau-Lifschitz system and spontaneous stochasticity would then manifest as non-trivial measures 
$P_{\bu_0}^{(t)}$ for $t>0$ with appropriate deterministic initial data $\mu=\delta_{\bu_0}.$ However,
this result would not immediately yield the conclusion on non-unique solutions of Euler equations 
without the knowledge that $P_\mu^{(t)}=\gamma_{t*}P_\mu$ for a space-time statistical solution 
$P_\mu$ of the type provided by our theorem. 
}
\end{rem} 

\begin{rem}\lb{rem1a}  
{\rm An essential feature of spontaneous stochasticity is the universality of the limiting measure $P_{\bu_0}$ which should be independent of the particular subsequence $\nu\to 0$ and in fact independent 
of the particular type of vanishing noise and regularization (which need not arise from thermal 
fluctuations or molecular viscosity). 
We obtain existence of a limit measure from tightness alone, so that we can guarantee convergence only along suitable subsequences, whose limits could in principle differ between subsequences. Previous rigorous examples of universal spontaneous statistics in model problems \cite{mailybaev2023spontaneous,mailybaev2023spontaneously} have relied on very strong dynamical mixing properties at all scales, which imply in particular a broad spectrum of positive Lyapunov exponents tending to infinite values as $\nu\to 0.$ Since current mathematical methods for realistic systems like the Landau-Lifschitz equations \eqref{llns} yield only a positive lower bound for the leading Lyapunov exponent \cite{bedrossian2022regularity,bedrossian2024chaos}, 
it does not seem feasible presently to prove the uniqueness or ``universality'' of the limit measures whose existence we establish. See \cite{mailybaev2024rg} for a numerical RG approach.} 
\end{rem} 

\begin{rem}\lb{rem2} 
{\rm We conjecture that the limit probability measures of the theorem are supported not only on weak Euler solutions 
but in fact on solutions which are locally dissipative. If true, then this would mean that the non-uniqueness
of dissipative Euler solutions proved by convex integration methods \cite{de2010admissibility,de2017high,daneri2021non}
is in fact physically realizable. Unlike for deterministic Navier-Stokes solutions \cite{duchon2000inertial}, it is not 
so elementary to prove for the Landau-Lifschitz equations that the weak Euler solutions obtained in the infinite-$Re$ limit 
are dissipative. The essential difficulty can be understood by considering the local kinetic energy balance for 
the Landau-Lifschitz equations, which we state here for equation \eqref{FNS2} written in dimensional variables:}
\bea 
&& \mathrm{d}\left(\frac{1}{2}|\bu|^2\right)+
\grad\cdot\left[\left(\frac{1}{2}|\bu|^2+p\right)\bu-\nu_\Lambda\grad\left(\frac{1}{2}|\bu|^2\right)\right] \mathrm{d}t \cr 
&& \hspace{50pt} =-\nu_\Lambda|\grad\bu|^2 \mathrm{d}t+ \sqrt{\frac{2\nu_\Lambda k_B T}{\rho}}\bu\bdot (\grad\cdot \mathrm{d}\bxx) +\frac{2\nu_\Lambda k_BT}{\rho} \cdot
\frac{1}{L^3} \sum_{|\bk|<\Lambda} |\bk|^2 \mathrm{d}t. \eea
{\rm Here $L$ is the side-length of the periodic flow domain. For the derivation, see e.g. \cite{flandoli2008introduction}. 
Note that this equation reduces only approximately to the energy balance for deterministic Navier-Stokes, if one 
chooses the wavenumber cutoff $\Lambda$ in a broad range $\gtrsim 1/\eta,$ with $\eta$ the Kolmogorov length, and if 
one considers only typical value of the thermal noise term proportional to $\bxx.$ See \cite{eyink2007turbulenceI}, 
Chapter II.E, for more details. It seems likely to us that the limit measures of our theorem can be proved 
to be supported on dissipative Euler solutions by using large deviations estimates similar to those of \cite{gess2023landau}
(although note that these authors do not consider the infinite-$Re$ limit, but instead treat the ``hydrodynamic scaling limit'' \cite{bardos1991fluid,bardos1993fluid,quastel1998lattice},
which was shown in \cite{bandak2022dissipation,bell_nonaka_garcia_eyink_2022} to be unphysical.) 
It is reasonable to expect, 
however, that if one considers not the $Re\to\infty$ limit itself, but instead the large deviations from that limit, then 
non-dissipative Euler solutions will be relevant. In fact, the hyperbolic scaling limit of the TASEP (totally asymmetric exclusion 
process) that leads to the inviscid Burgers equation is quite similar to the infinite Reynolds limit, and there the 
non-dissipative weak solutions of inviscid Burgers equation govern the large deviations probabilities 
\cite{jensen2000large,varadhan2004large,quastel2021hydrodynamic}.}  
\end{rem} 

\begin{rem}\lb{rem3}
{\rm We conjecture also that the limiting probability measures of our theorem give the path distributions 
of a Markov process in time. In fact, in all rigorous examples of which we are aware, the limiting 
spontaneously stochastic process is Markov
\cite{lejan2002integration,lejan2004flows,eyink2015spontaneous,mailybaev2023spontaneous,mailybaev2023spontaneously}. 
Furthermore, the Landau-Lischitz dynamics \eqref{llns} for $\nu>0$ is proved to be Markov by standard 
arguments \cite{flandoli2008introduction} and this property is expected to be preserved in the limit 
$\nu\to 0$ under reasonable assumptions \cite{karr1975weak}.}  
\end{rem} 

\begin{rem}\lb{rem4} 
{\rm A comment should be made about the regularity assumptions on our initial data $\bu_0.$ As discussed earlier,
it is convenient to choose initial data that are non-smooth, so that spontaneous stochasticity appears 
immediately, rather than after some finite time sufficient, hypothetically, for the solution to blow-up. 
However, the regularity $\bu_0\in L^p({\mathbb T}^d)$ is much rougher than required, and it would be much 
more natural physically to take $\bu_0\in B^\sigma_{p,\infty}({\mathbb T}^d)$ for some $p>2$ and $0<\sigma<1.$
We did not do so, because our proof of part (iv) of the theorem requires separability of $L^p({\mathbb T}^d),$
whereas $B^\sigma_{p,\infty}({\mathbb T}^d)$ is non-separable. Two comments are useful. First, there is reason 
to believe that dissipative weak solutions of the Euler equations will have some spatial Besov regularity 
immediately at any time $t>0,$ even if the initial data at $t=0$ is rougher \cite{drivas2022self}. Second,
we have checked that our theorem remains true if we employ the Besov space $B^\sigma_{p,q}({\mathbb T}^d)$
for $q<\infty$ rather than for $q=\infty,$ as stated above. Since the Besov spaces are separable for 
$p,q<\infty,$ we can in that case choose initial data $\bu_0\in B^\sigma_{p,q}({\mathbb T}^d)$ and 
our current technique of proof applies. Furthermore, part (ii) of the theorem can then be strengthened by 
replacing $\widetilde{\mathcal{E}}^{\sigma,\beta}_{r,p}$ instead with 
$$\mathcal{E}^{\sigma,\beta,\gamma}_{r,p,q}=L^r([0,T],B_{p,q}^{\sigma}(\mathbb{T}^d))
\cap C^{0,\gamma}([0,T],B_{p/2,q/2}^{\sigma-\beta}(\mathbb{T}^d))$$
which supplies the realizations of the limit measure with some H\"older continuity in time.}  
\end{rem}

\begin{rem}\lb{rem5}
{\rm A similar comment is required about the assumed regularity of the force $\bsf$ in our theorem.
In fact, for parts (i),(ii) the proofs go through easily assuming only $\bsf\in C^0([0,T],B_{p/2,\infty}^{\sigma-\beta}(\mathbb{T}^d))$ 
rather than our more stringent assumption $\bsf\in C^0([0,T]\times \mathbb{T}^d).$ It seems possible that, with some further 
work, we could mathematically strengthen our theorem to require only $\bsf\in C^0([0,T],B_{p/2,\infty}^{\sigma-\beta}(\mathbb{T}^d))$ 
for all of its statements. However, even the condition $\bsf\in C^0([0,T]\times \mathbb{T}^d)$ has dubious physical relevance.
To be clear, a physical forcing should represent stirring only at large scales, so that the force should be extremely 
smooth in space, e.g. $\bsf\in C^0([0,T],{\mathcal S}(\mathbb{T}^d)).$ Besov regularity of the
type \eqref{besov} that we have hypothesized is very unlikely to hold when the fluid is forced with 
increasing strength at decreasing length scales.}  
\end{rem}

\begin{rem}\lb{rem6} 
{\rm Finally, we make a comment about the dependence of our results on the space dimensionality $d.$
The non-dimensionalization of the equation \eqref{FNS3} assumes for any space dimension $d$ the relation $\varepsilon\sim U^3/L$ of Taylor \cite{taylor1935statistical}, which is an expression of anomalous dissipation 
and energy cascade. See section \ref{sec:back}. 
In fact, for $d>3$ there are both theoretical arguments and numerical evidence  
\cite{suzuki2005energy,gotoh2007statistical,yamamoto2012local,berera2020homogeneous} that a turbulent energy cascade 
to small scales exists, just as in $d=3.$ If the Taylor relation is, in fact, modified by an $Re$-dependent 
factor, then the exponent $\kappa=3(d+2)/8$ of the $\nu$-dependent prefactor in the thermal noise will be slightly 
modified but remain positive. However, as is well-known, turbulence phenomenology is quite different 
for $d=2,$ where there is no energy cascade to small scales but instead an enstrophy cascade \cite{boffetta2012two}.
In addition, the consequences of thermal noise are much more profound for $d=2,$ where theory predicts that 
the renormalized viscosity $\nu_\Lambda$ should depend logarithmically both upon the wavenumber cutoff 
$\Lambda$ and the flow domain size $L$ \cite{forster1977large}. The logarithmic divergence with $L$ has been observed
in some lattice-gas simulations of $d=2$ laminar shear flow \cite{kadanoff1989automata}. We conjecture 
that this logarithmic divergence in a forced turbulent flow in $d=2$ will depend not upon $L$ but instead 
upon the Kraichnan dissipation scale that terminates the enstrophy cascade range. Although we have stated 
our mathematical theorem for $d\geq 2,$ many of our assumptions and results must be modified 
to have physical relevance for $d=2.$} 
\end{rem} 

\section{Theoretical and Empirical Background}\lb{sec:back} 

In this section, prior to proving our main theorem, we give some necessary background 
on Landau-Lifschitz fluctuating hydrodynamics, its physical justification, empirical 
verification and rigorous status. We shall discuss also the non-dimensionalization 
of the Landau-Lifshitz equation with large-scale flow parameters that underlies our theorem 
and the empirical basis of the uniform Besov regularity which is its main hypothesis. 
Readers who are interested mainly in the mathematical results can skip this section on 
a first reading and go directly to section \ref{sec:proofs} where the proofs are presented. 

Fluctuating hydrodynamics was originally proposed on a phenomenological basis by Landau \& Lifschitz \cite{landau1959fluid}, who added a fluctuating stress with statistics governed by the fluctuation-dissipation 
relation \eqref{FDR}  to the fluid equations, as the simplest hydrodynamic model to reproduce 
the known Boltzmann-Einstein single-time statistics in local thermodynamic equilibrium. 
It was later pointed out in \cite{vansaarloos1982non} that the original Landau-Lifschitz 
theory reproduced the correct local equilibrium fluctuations only if transport coefficients are 
state-independent. This observation triggered efforts to derive fluctuating hydrodynamics 
microscopically by formal methods of statistical physics \cite{zubarev1983statistical},
with the conclusion that the correct noise given by the fluctuation-dissipation relation must be multiplicative \cite{morozov1984langevin}. See also 
\cite{espanol1998stochastic,espanol2009microscopic}. We are aware of no mathematically 
rigorous derivation of the Landau \& Lifschitz equations for nonlinear fluctuating hydrodynamics, 
as necessary to describe turbulent flow. The closest result of which we are aware is 
a rigorous derivation of a large deviations rate function for a microscopic lattice gas in 
a ``hydrodynamic scaling limit'' at small Knudsen and Mach numbers but for arbitrary
(fixed) Reynolds number \cite{quastel1998lattice}. The large-deviations rate function
obtained in this work corresponds formally to that obtained from the incompressible 
Landau-Lifschitz theory in a weak-noise limit \cite{eyink1990dissipation,gess2023landau}. Unfortunately, 
subsequent work \cite{bandak2022dissipation,bell_nonaka_garcia_eyink_2022} has shown 
that the assumed ``hydrodynamic scaling limit'' is physically unrealistic for turbulent 
flows and probably never applies anywhere in Nature. 
On the other hand, the Landau-Lifschitz theory 
predicts non-trivial effects of thermal noise, especially in flow states out of global
thermodynamic equilibrium, which are extensively confirmed by laboratory experiments 
for many laminar flows \cite{dezarate2006hydrodynamic}. Since the main physical
assumption of the Landau-Lifschitz theory is a local equilibrium property, it is reasonable 
to expect that fluctuating hydrodynamics applies also to turbulent flows as long as this 
fundamental property continues to hold there. 

A basic conclusion of physicists' microscopic derivations \cite{zubarev1983statistical,espanol2009microscopic} 
of nonlinear Landau-Lifschitz equations is that, despite superficial appearances, they are {\it not} 
stochastic partial differential equations. Instead, these equations contain a specific high-wavenumber 
cutoff $\Lambda$ and are designed to describe only the length scales $>\Lambda^{-1}.$ In the language 
of physics, the Landau-Lifschitz hydrodynamics equations are low-wavenumber ``effective field theories"
\cite{schwenk2012renormalization,liu2018lectures}. A standard feature of such theories is that the 
cutoff $\Lambda$ is arbitrary 
to a large degree but that the parameters of the model depend upon $\Lambda$ or ``run'' with $\Lambda$
according to renormalization-group flow equations, so that the objective predictions of the model
are $\Lambda$-independent. All of these features are nicely illustrated by the incompressible model 
\eqref{FNS2}. Here we may note that this model follows from the general Landau-Lifshitz theory for a 
simple Newtonian fluid in the limit of small Mach numbers, as discussed carefully in \cite{donev2014low}. 
A renormalization group analysis of Forster, Nelson \& Stephen \cite{forster1977large} studied
the $\Lambda$-dependence of the viscosity $\nu_\Lambda$ in the model \eqref{FNS2} for a fluid in 
global thermal equilibrium (no mean motion) with the well-known energy spectrum 
$E(k)\propto \frac{k_B T}{\rho} k^2$ of thermal velocity fluctuations at absolute temperature $T$.  
They found for $d=3$ (3D) that the renormalization of the viscosity by thermal fluctuations 
is rather weak and of the form $\nu_\Lambda = \nu - O(\Lambda k_BT/\rho \nu)$ where $\nu$ is the measured
or ``macroscopic'' viscosity for very small $\Lambda.$ These same results should hold also for 3D turbulent 
flows if one takes $\Lambda>1/\eta,$ where $\eta=\nu^{3/4}/\varepsilon^{1/4}$ is the Kolmogorov dissipation 
length defined in terms of the macroscopic viscosity $\nu$ and the mean rate $\varepsilon$ of 
kinetic energy injection by the external force (per unit time and per unit mass). 
Indeed, numerical simulations of 3D turbulence which include thermal fluctuations show 
that the energy spectrum $E(k)\propto \frac{k_B T}{\rho} k^2$ of these fluctuations 
become dominant just below the Kolmogorov length 
\cite{bandak2022dissipation,bell_nonaka_garcia_eyink_2022,mcmullen2022navier}. In that case,
any choice of $\Lambda^{-1}$ below the Kolmogorov length $\eta$ and well above the mean-free-path
length $\lambda_{mfp}$ of the fluid should suffice to yield \eqref{FNS2} as a valid model
of turbulent flow, with roughly the same choice of $\nu_\Lambda$ as in thermal 
equilibrium. Note, however, that this model might break down locally in spacetime due to 
extreme turbulent fluctuations at sufficiently high $Re$ \cite{bandak2022dissipation}. 
Although the model \eqref{FNS2} has solutions which are mathematically 
well-defined globally in time \cite{flandoli2008introduction}, the solutions might develop 
extremely energetic small scales $\sim \Lambda^{-1}$ which invalidate the physical derivations 
\cite{zubarev1983statistical,espanol2009microscopic}.  

An important application of Theorem \ref{result2} is to solutions of the Landau-Lifschitz 
equations \eqref{FNS2} that appear as typical realizations selected from a turbulent flow. 
Therefore its statement is formulated 
in terms of the equations \eqref{FNS3} non-dimensionalized in the standard manner for the 
inertial scales of high-$Re$ turbulence. As usual, one introduces dimensionless variables
$$ \hat{{\bf x}}={\bf x}/L, \quad \hat{t}=t/(L/U), \quad \hat{{\bf u}}={\bf u}/U,\quad
\hat{p}=p/U^2 $$
where $L$ is the outer or stirring length (defined by the scale of the forcing ${\bf f}$)
and $U$ is the velocity of the energy-containing eddies (for example, the rms turbulent
velocity $u_{rms}$). After introducing these rescalings and omitting the hats, for convenience, 
the equation \eqref{FNS3} is obtained. In performing the non-dimensionalization, we have 
furthermore assumed Taylor's relation $\varepsilon \sim U^3/L$ \cite{taylor1935statistical}, 
sometimes called the ``zeroth-law of turbulence.'' Although there is considerable empirical 
support for this relation \cite{sreenivasan1984scaling,kaneda2003energy}, it is possible
that $\hat{\varepsilon}=\varepsilon/(U^3/L)$ vanishes slowly with $Re$ 
\cite{drivas2019onsager,bedrossian2019sufficient,iyer2023dependence,iyer2024zeroth} and this could 
change the $Re$-dependencies slightly, but without vitiating our analysis.  
The dimensionless number groups that appear in \eqref{FNS3} in addition to $Re$ are 
the parameters $\theta_\eta=k_BT/\rho u_\eta^2\eta^d$ (where $u_\eta=(\varepsilon\nu)^{1/4}$ 
is the Kolmogorov velocity) and $\digamma=Lf_{rms}/U^2$ (where $f_{rms}$ is the rms amplitude
of the external force). The infinite-$Re$ limit is one with $UL\gg \nu$ due to large $L$ and $U$
\footnote{The incompressibility approximation will eventually break down in this limit  
when $U=(\varepsilon L)^{1/3}$ approaches $c_s$, the sound speed. The underlying assumption in using
model \eqref{FNS2} is that the infinite-$Re$ asymptotics will be achieved before compressibility 
effects appear at large scales. Incompressibility may also break down at the UV end if $\Lambda^{-1}$ 
is chosen sufficiently close to the interparticle distance $n^{-1/3}$ so that the thermal velocity 
fluctuation $\sim c_s (\Lambda^3/n)^{1/2}$ is an appreciable fraction of the sound speed.} 
and fixed fluid parameters such as $\rho,$ $T,$ and $\nu,$ so that $\theta_\eta$ is held fixed. 
With Taylor's relation $\digamma\sim f_{rms}U/\varepsilon$ measures the ``inefficiency'' of the force
for energy input and will be assumed fixed as well.
Finally, $\Lambda$ is kept fixed, but note that $\hat{\Lambda}\to\infty.$
For example, $\hat{\Lambda}\sim Re^{3/4}$ if $\eta\Lambda$ is held fixed, or, using the 
standard kinetic theory result $\nu\sim c_s\lambda_{mfp},$ then instead $\hat{\Lambda}\sim Re$ if 
$\lambda_{mfp}\Lambda$ is held fixed. 

We consider finally the empirical basis of the {\it a priori} Besov regularity that is assumed 
in Theorem \ref{result2}. The primary focus of interest is a measure ${\mathbb P}_*$ on spacetime
velocity fields, e.g. a statistically stationary turbulent solution with time-independent 
single-instant marginals or perhaps a decaying turbulent solution with time-dependent marginals. 
It was shown already some time ago by a simple Borel-Cantelli argument \cite{eyink1995besov} 
that power-law scaling (or even upper bounds) of ensemble-average absolute structure functions
\begin{equation} 
S_p(r):={\mathbb E}_*\left[|\delta {\bf u}({\bf r})|^p\right]\sim 
C_p U^p\left(\frac{r}{L}\right)^{\zeta_p} \lb{Sp-scal} \end{equation}
obtained in the limit $Re\to\infty,$ together with ${\mathbb E}_*|{\bf u}|^p<\infty,$
implies that $\bu(\cdot,\omega)\in B^{\sigma}_{p,\infty}({\mathbb T}^d)$ for any choice 
$\sigma<\sigma_p:=\zeta_p/p$ and for ${\mathbb P}_*$-a.e. sample point $\omega.$ In that case,  
for any $\epsilon>0$
\begin{equation}
\|\delta{\bf u}({\bf r})\|_p^p:=
\frac{1}{V}\int_{{\mathbb T}^d} |\delta {\bf u}({{\bf x};\bf r})|^p \,\mathrm{d}^dx
\leq C_p(\omega) U^p\left(\frac{r}{L}\right)^{\zeta_p-p\epsilon},
\qquad {\mathbb P}_*\mbox{-a.e. } \omega \label{Sp-ineq} \end{equation}
with the optimal constant given by the Besov semi-norm 
\begin{equation}
    C_p(\omega)=\vvvert {\bf u}(\cdot,\omega)\vvvert^p_{B^{\sigma_p-\epsilon}_{p,\infty}}
    :=\sup_{r<L} \frac{\|\delta{\bf u}({\bf r})\|_p^p}{U^p\left(r/L\right)^{\zeta_p-p\epsilon}},
\qquad {\mathbb P}_*\mbox{-a.e. } \omega. 
\label{Cp-def} \end{equation}    
These considerations apply at each time instant, implying that \eqref{Sp-ineq}
holds for increments $\delta {\bf u}({{\bf x},t;\bf r})$ at each time $t\in [0,T]$
with the constant $C_p(\omega,t)=\vvvert {\bf u}(\cdot,t,\omega)
 \vvvert^p_{B^{\sigma_p-\epsilon}_{p,\infty}}.$ In that case, the reasonable physical assumptions that 
\begin{equation}  
\frac{1}{T}\int_0^T{\mathbb E}_*[|{\bf u}(\cdot,t)|^p]\,\mathrm{d}t<\infty, \quad 
\frac{1}{T}\int_0^T {\mathbb E}_*[C_p(t)]\,\mathrm{d}t<\infty
\label{besov-cond} 
\end{equation}
are equivalent to the mathematical statement that ${\mathbb E}_*\left[\|{\bf u}\|^p_{L^p([0,T],
B^{\sigma_p-\epsilon}_{p,\infty})}\right]<\infty.$ The above considerations for the ideal limit
$Re=\infty$ are easily extended to the realistic case of large but finite $Re.$ Indeed, 
with ${\mathbb E}_*$ replaced by ${\mathbb E}^\nu_*,$ the relation 
\eqref{Sp-scal} is observed to hold both in laboratory experiments and in numerical simulations 
but only over an ``inertial range'' of length-scales $\eta_p<r<L.$ For example, see  
\cite{anselmet1984high,chen2005anomalous,iyer2020scaling}. From simulations,  
$S_p^\nu(r)$ takes on much smaller values $C_p' \langle|{\mathbb \grad}{\bf u}|^p\rangle r^p$
in the ``dissipation range'' $r<\eta_p$.
In that case, \eqref{Sp-ineq},\eqref{Cp-def} still hold for each $\nu=1/Re$ and 
${\mathbb P}^\nu_*$-a.e. $\omega,$ with the supremum in \eqref{Cp-def} achieved 
in a range $\eta_p(\omega)<r<L.$ Thus, the conditions \eqref{besov-cond} are plausibly 
assumed to hold uniformly in $\nu=1/Re$ so that $\sup_{\nu>o}{\mathbb E}^\nu_*\left[\|{\bf u}\|^p_{L^p([0,T],
B^{\sigma_p-\epsilon}_{p,\infty})}\right]<\infty,$ which for $p>2$ is stronger than the 
fundamental assumption \ref{besov} of Theorem \ref{result2}. 

The previous considerations do not explicitly include thermal noise. Physical experiments
do include all molecular effects implicitly, of course, but there are no observations
at sub-Kolmogorov scales, so that only the inertial-range scaling \eqref{Sp-scal}
can be confirmed in a range $\eta_p\ll r\ll L$ \cite{anselmet1984high,chen2005anomalous,iyer2020scaling}
and near-dissipation range drop-off 
for $r\sim \eta_p$ \cite{buaria2020dissipation,gorbunova2020analysis}. 
Since velocity fluctuations from thermal 
noise will increase with the wavenumber cutoff $\Lambda$ and eventually will even exceed 
the low-Mach turbulent fluctuations, there is a serious concern that our uniform {\it a priori}  
bounds will be violated. The only empirical source of information about the effects of thermal 
noise on turbulence come from the few numerical simulations that have so far included them
\cite{bell_nonaka_garcia_eyink_2022,mcmullen2022navier,mcmullen2023thermal}. Out of this 
small set of studies, only \cite{mcmullen2023thermal} so far has studied the effects 
of thermal noise on turbulent structure functions, using the Direct Simulation Monte Carlo 
(DSMC) method for $d=3$. Most useful for us, \cite{mcmullen2023thermal} found that their numerical 
results are well-described by a simple ``superposition model'', according to which 
the velocity coarse-grained over length-scales $<\Lambda^{-1}$ can be represented as
\begin{equation}
{\bf u }({\bf x},t) = {\bf u }^{NS}({\bf x},t) + \left(\frac{Fk_BT}{\rho}\Lambda^3\right)^{1/2}
{\bf N}({\bf x},t)
\end{equation}
where ${\bf u }^{NS}({\bf x},t)$ is the deterministic Navier-Stokes solution and ${\bf N}({\bf x},t)$
is a standard normal random variable. The latter is chosen independently for every discrete value 
of ${\bf x}$ and $t$ in the DSMC simulation and independent of the Navier-Stokes solution. 
In the DSMC method the quantity $F$ 
is the ``simulation ratio,'' a large factor which prescribes  
the number of true molecules represented by each DSMC computational molecule, but in the 
physically realistic limit $F$ is a number of order unity. This model yields the following 
predictions for the first few even-order velocity structure functions at separations
$\Lambda r>1$:
\begin{eqnarray}
S_2(r) &=& S_2^{NS}(r) + 2 \frac{Fk_BT}{\rho}\Lambda^3   
\label{S2} \end{eqnarray}
\begin{eqnarray}
S_4(r) &=& S_4^{NS}(r) + 12 \left(\frac{Fk_BT}{\rho}\Lambda^3\right) S_2^{NS}(r)  
+ 12 \left(\frac{Fk_BT}{\rho}\Lambda^3\right)^2
\label{S4} \end{eqnarray}
\begin{eqnarray}
S_6(r) &=& S_6^{NS}(r) + 30 \left(\frac{Fk_BT}{\rho}\Lambda^3\right) S_4^{NS}(r)  \cr
&& \hspace{10pt} + 180 \left(\frac{Fk_BT}{\rho}\Lambda^3\right)^2 S_2^{NS}(r) 
+ 120 \left(\frac{Fk_BT}{\rho}\Lambda^3\right)^3
\label{S6} \end{eqnarray}
See \cite{mcmullen2023thermal}, Eq.(6). Note that the superposition model becomes  
open to question if one considers structure functions of higher order $p\gg 1,$ 
since these quantities are dominated by very rare, extreme turbulent fluctuations 
which might penetrate to very tiny length scales even smaller than $\Lambda^{-1}$ 
and thus disrupt the assumption of local equilibrium at scale $\Lambda$ 
\cite{bandak2022dissipation}. We employ the model here as the best available 
representation of the effects of thermal noise on structure functions of low orders, 
conveniently summarizing the numerical observations in \cite{mcmullen2023thermal}
and also the related observations for a shell model in \cite{bandak2022dissipation}. 
Finally, we note that for distances $\Lambda r<1$ the structure functions decrease
rapidly below the above values, with $S_p(r)=O(|\Lambda r|^p),$ since the velocity 
field within the ``effective field theory'' is band-limited to wavenumbers $<\Lambda.$

To carry over our previous argument about Besov regularity of velocity realizations,
we must show that the scaling relation \eqref{Sp-scal} survives for 
{\it all} $r<L$ in the presence of thermal fluctuations, at least in the modest sense 
of an upper bound. Thus, essentially, we must establish finiteness of the following 
constant:
\begin{equation} 
C_p:=\sup_{r<L}\frac{S_p(r)}{U^p\left(r/L\right)^{\zeta_p}}<\infty.    
\end{equation}
We naturally assume finiteness of the corresponding constant for the deterministic 
Navier-Stokes solution: $C_p^{NS}=\sup_{r<L}\frac{S^{NS}_p(r)}{U^p\left(r/L\right)^{\zeta_p}}<\infty.$
To estimate the thermal contribution we must consider the key quantity
\begin{eqnarray}
    C_p^{th}(Re)&:=& \sup_{\Lambda^{-1}<r<L} \left(\frac{Fk_BT}{\rho}\Lambda^3\right)^{p/2} 
        \frac{1}{U^p}\left(\frac{L}{r}\right)^{\zeta_p} \cr 
   &=&   \left(\frac{Fk_BT}{\rho U^2}\Lambda^3\right)^{p/2} (L\Lambda)^{\zeta_p}
  \, =\,   \Theta_\eta^{p/2} \left(\frac{\eta}{L}\right)^{\frac{p}{3}-\zeta_p} 
       (\eta\Lambda)^{\frac{3p}{2}+\zeta_p}. 
\lb{Cpth} \end{eqnarray}       
Notice, because of the rapid decay of $S_p(r)$ for $\Lambda r<1,$ that the supremum 
over all $r<L$ is in fact achieved for $\Lambda r>1$ and this leads to the restriction 
of the range in \eqref{Cpth}. We have also defined $\Theta_\eta:=F\theta_\eta.$
In terms of the two constants $C_p^{NS}$ and $C_p^{th}$
we immediately obtain from \eqref{S2}
\begin{equation}
C_2=\sup_{\Lambda^{-1}<r<L}\frac{S_2(r)}{U^2\left(r/L\right)^{\zeta_2}}
  \leq C_2^{NS} + 2 C_2^{th},  
\label{C2} \end{equation}  
from \eqref{S4} using the H\"older inequality $S_2^{NS}(r)\leq [S_4^{NS}(r)]^{1/2}$
\begin{equation}
C_4=\sup_{\Lambda^{-1}<r<L}\frac{S_4(r)}{U^4\left(r/L\right)^{\zeta_4}}
  \leq C_4^{NS} + 12 (C_4^{th})^{1/2}(C_4^{NS})^{1/2}+ 12 C_4^{th} 
\label{C4} \end{equation}    
from \eqref{S6} using  H\"older inequalities $S_4^{NS}(r)\leq [S_6^{NS}(r)]^{2/3},$ $S_2^{NS}(r)\leq [S_6^{NS}(r)]^{1/3},$ 
\begin{equation}
\sup_{\Lambda^{-1}<r<L}\frac{S_6(r)}{U^6\left(r/L\right)^{\zeta_6}}
  \leq C_6^{NS} + 30 (C_6^{th})^{1/3} (C_6^{NS})^{2/3} +180 (C_6^{th})^{2/3}(C_6^{NS})^{1/3}
  + 120 C_6^{th}, 
\label{C6} \end{equation}    
and so forth. Clearly, the behavior of $C_p^{th}(Re)$ for $Re\to\infty$ is crucial. 

To complete the argument we must appeal to some properties of the turbulent 
scaling exponents $\zeta_p,$ which are known to be concave in $p$ and which coincide with the 
K41 line $p/3$ at exactly two $p$-values, $p=0$ and $p=p_*\doteq 3$
\cite{frisch1995turbulence}. Note that, because we deal here with structure 
functions of absolute velocity increments, the Kolmogorov 4/5th law does not apply,  
so that it is not true that $p_*=3$ exactly. The important consequence is that 
$\zeta_p<p/3$ for $p>p_*$ and we can thus infer from the bound \eqref{Cpth} that 
$$ \lim_{Re\to\infty} C_p^{th}(Re)=0,\qquad p>p_*, $$
{\it if one assumes that $\eta\Lambda$ is held fixed}. In conjunction with the estimates \eqref{C4} for $C_4,$ \eqref{C6} for $C_6,$ etc. we see that all of these constants remain bounded in the presence of thermal noise and, indeed, coincide for $Re\gg 1$ with the Navier-Stokes values $C_4^{NS},$$C_6^{NS},$ etc. The Besov regularity of the velocity realizations is thus unchanged from the deterministic case
when the cutoff $\Lambda$ is taken to be some fixed multiple of $\eta^{-1}.$ In fact, one
can also take $\eta\Lambda$ increasing with $Re$ at a sufficiently slow rate, since the 
only requirement is that $C_p^{th}(Re)$ remain bounded as $Re\to\infty.$ In that case,
\eqref{Cpth} implies that the choice
\be \eta\Lambda\, \propto \left(\frac{L}{\eta}\right)^{\frac{\frac{p}{3}-\zeta_p}{\frac{3p}{2}+\zeta_p}}
\sim Re^{\frac{p-3\zeta_p}{6p+4\zeta_p}}
\ee 
would suffice to guarantee that $C_p(Re)$ remain bounded as $Re\to\infty,$ although with a 
different limiting value of the constant than for the deterministic case. With either choice 
of $\Lambda$ one can conclude that 
$$\sup_{\nu>o}{\mathbb E}^\nu_*\left[\|{\bf u}\|^p_{L^p([0,T],
B^{\sigma_p-\epsilon}_{p,\infty})}\right]<\infty,\qquad  p=4,6,8,... $$
at least for even $p$ that are not too large.  

On the other hand, for $p<p_*$ instead $\zeta_p>p/3,$
and thus $\lim_{Re\to\infty}C_p^{th}(Re)=\infty!$ In that case, one can instead 
replace $\zeta_p$ with $p/3$ to obtain 
$$ \widetilde{C}_p^{th}(Re):= \sup_{\Lambda^{-1}<r<L} \left(\frac{Fk_BT}{\rho}\Lambda^3\right)^{p/2} 
        \frac{1}{U^p}\left(\frac{L}{r}\right)^{p/3}
  =   \Theta_\eta^{p/2} 
       (\eta\Lambda)^{\frac{11p}{6}}<\infty, $$
whereas also $\widetilde{C}_p^{NS}=\sup_{r<L}\frac{S^{NS}_p(r)}{U^p\left(r/L\right)^{p/3}}<\infty.$
Thus, choosing the cutoff $\Lambda$ once again to be some fixed multiple of $\eta^{-1},$ one can obtain 
at least K41 regularity of the velocity field for $p<p_*.$ In particular, using \eqref{C2} for $p=2$ 
it is justified by the empirical observations to assume that
$$\sup_{\nu>o}{\mathbb E}^\nu_*\left[\|{\bf u}\|^2_{L^2([0,T],
B^{\frac{1}{3}-\epsilon}_{2,\infty})}\right]<\infty. $$

\section{Proof of the Main Theorem}\lb{sec:proofs}  

We follow the general strategy of \cite{vishik1988mathematical}, Chapter IV, who proved the existence of space-time statistical 
solutions of the incompressible Navier-Stokes equations, thereby obtaining a probabilistic version of Leray's construction. 
We give the proofs of parts (i)-(iv) of our main theorem in successive subsections. We use frequently 
the convenient notation $a\lesssim b$ with the meaning that $a\leq C b$ for some constant $C$
depending possibly upon parameters that are fixed. 

\subsection{Proof of (i):} Our proof of tightness is modeled after Theorem IV.4.3 of \cite{vishik1988mathematical} and is likewise based upon their following general result: 

\vspace{10pt} 
\noindent 
{\bf Lemma II.3.1 of \cite{vishik1988mathematical}}: {\it Let $\Omega_1,$ $\Omega_2$ be Banach spaces, where $\Omega_2$ is 
separable and $\Omega_1$ is compactly embedded in $\Omega_2.$ The family $\mathscr{M}$ of probability measures defined
on ${\mathcal B}(\Omega_2)$ and with support on $\Omega_1$ is tight on ${\mathcal B}(\Omega_2)$ if for any 
$\mu\in \mathscr{M},$ the functional $u\mapsto \|u\|_{\Omega_1}$ is $\mu$-integrable and 
\be \sup_{\mu\in\mathscr{M}} \int \|u\|_{\Omega_1} d\mu(u) <\infty \lb{tight-bd} \ee}
\vspace{10pt} 
We apply this proposition with $\Omega_2=\mathcal{Z}^{\sigma,\beta}_{r,p}$ and with Banach space $\Omega_1=\mathcal{E}^{\sigma,\beta,\gamma}_{r,p}$ defined for 
$0<\gamma<1$ by
\[ \mathcal{E}^{\sigma,\beta,\gamma}_{r,p}:= L^r([0,T],B_{p,\infty}^{\sigma}(\mathbb{T}^d)) \cap C^{0,\gamma}([0,T],B_{p/2,\infty}^{\sigma-\beta}(\mathbb{T}^d))\]
where 
\[\lVert\cdot\rVert_{\mathcal{E}^{\sigma,\beta,\gamma}_{r,p}}:=\lVert\cdot\rVert_{L^r([0,T],B_{p,\infty}^{\sigma}(\mathbb{T}^d))}+\lVert\cdot\rVert_{C^{0,\gamma}([0,T],B_{p/2,\infty}^{\sigma-\beta}(\mathbb{T}^d))}.\]
First, we note that $\mathcal{Z}^{\sigma,\beta}_{r,p}\subseteq L^r([0,T],L^p(\mathbb{T}^d)),$ and the latter is a separable Banach space. By Proposition 3.25 of \cite{brezis2010functional}, we conclude $\mathcal{Z}^{\sigma,\beta}_{r,p}=L^r([0,T],L^p(\mathbb{T}^d))\cap C^0([0,T],B_{p/2,\infty}^{\sigma-\beta}(\mathbb{T}^d))$ 
is separable. 

Next we argue that
\[\mathcal{E}^{\sigma,\beta,\gamma}_{r,p}\hookrightarrow \mathcal{Z}^{\sigma,\beta}_{r,p}\]
is a compact embedding for $0<\gamma<\min((r-2)/r,1/2).$ We use the following version of the Aubin-Lions-Simons compactness result:

\vspace{10pt} 
\noindent 
{\bf Theorem IV.4.1 of \cite{vishik1988mathematical}}: {\it Let $E_0,$ $E,$ $E_1$ be Banach spaces, such that $E$ is 
continuously embedded in $E_1$ and $E_0$ is compactly embedded in $E.$ Let $1<q<\infty$ and $M$ be a bounded set 
in $L^q([0,T],E_0)$ consisting of functions equicontinuous in $C^0([0,T],E_1).$ Then $M$ is relatively compact
in $L^q([0,T],E)\cap C^0([0,T],E_1).$}

\vspace{10pt} 
\noindent 
We note that the original statement of this result assumed that all of the Banach spaces $E_0,$ $E,$ $E_1$ are reflexive. However,  we have checked 
that the proof of Theorem IV.4.1in \cite{vishik1988mathematical} used reflexivity only in their Lemma IV.4.1 and this result is now well-known 
not to require reflexivity. For example, see Lemma II.5.15 in section 5.3 of \cite{boyer2012mathematical}. We thus can apply the above proposition with $q=r$ and 
$$ E_0=B_{p,\infty}^{\sigma}(\mathbb{T}^d), \quad E=L^p(\mathbb{T}^d), \quad E_1=B_{p/2,\infty}^{\sigma-\beta}(\mathbb{T}^d)$$
even though $B_{p,\infty}^{\sigma}(\mathbb{T}^d)$ and $B_{p/2,\infty}^{\sigma-\beta}(\mathbb{T}^d)$ are not reflexive. The hypotheses 
of Theorem IV.4.1 in \cite{vishik1988mathematical} are verified in the following lemmas.

First the compact embedding of $E_0$ into $E$ is given by 
\begin{lemma}\label{claim1} For any $1\leq p<\infty$, $0<\sigma<1$, and $d\in\mathbb{N}$ the embedding $B_{p,\infty}^{\sigma}(\mathbb{T}^d)\hookrightarrow L^p(\mathbb{T}^d)$ is compact.\end{lemma}
\hfill
\textit{Proof.} Following the Kolmogorov--Riesz--Frech\'et theorem (Theorem 4.26 of \cite{brezis2010functional}) it suffices to argue that for any 
bounded subset $\Omega\subseteq B_{p,\infty}^{\sigma}(\mathbb{T}^d),$
\[\lim_{|\mathbf{h}|\to0}\sup_{\bvp\in\Omega}\left\lVert\bvp(\cdot+\mathbf{h})-\bvp(\cdot)\right\rVert_{L^p(\mathbb{T}^d)}=0.\]
However, by definition of $B_{p,\infty}^{\sigma}(\mathbb{T}^d),$
\[\sup_{\bvp\in\Omega}\left\lVert\bvp(\cdot+\mathbf{h})-\bvp(\cdot)\right\rVert_{L^p(\mathbb{T}^d)}\leq|\mathbf{h}|^\sigma\sup_{\bvp\in\Omega}\lVert\bvp\rVert_{B_{p,\infty}^{\sigma}(\mathbb{T}^d)}\to0 \ \mbox{ as } \ h\to 0. \qed \]

\hfill

Next, the continuous embedding of $E$ into $E_1$ follows from 
\begin{lemma}\label{claim2}For any $1\leq p'\leq p<\infty$, $0<\sigma<1$, $\beta\geq 2$ and $d\in\mathbb{N}^+$ the embedding $L^p(\mathbb{T}^d)\hookrightarrow B_{p',\infty}^{\sigma-\beta}(\mathbb{T}^d)$ is continuous.\end{lemma}

\hfill

\textit{Proof.} From the famous Littlewood--Paley theorem (Remark 2 of 3.5.4 in \cite{schmeisser1987topics}), we have equivalence of $L^p(\mathbb{T}^d)$ 
with $F_{p,2}^{0}(\mathbb{T}^d),$  the Triebel--Lizorkin space. Then we have a chain of continuous embeddings (Remark 4 of 3.5.1 in \cite{schmeisser1987topics})
\[F_{p,2}^{0}(\mathbb{T}^d)\hookrightarrow B_{p',2}^{0}(\mathbb{T}^d)\hookrightarrow B_{p',2}^{\sigma-\beta}(\mathbb{T}^d)\hookrightarrow B_{p',\infty}^{\sigma-\beta}(\mathbb{T}^d). \hfill \qed \]

\hfill 

Finally, boundedness in $C^{0,\gamma}([0,T],B_{p/2,\infty}^{\sigma-\beta}(\mathbb{T}^d))$ immediately implies equicontinuity from $[0,T]$ into $B_{p/2,\infty}^{\sigma-\beta}(\mathbb{T}^d)$. Thus, any bounded subset of $\mathcal{E}^{\sigma,\beta,\gamma}_{r,p}$ is relatively compact in $\mathcal{Z}^{\sigma,\beta}_{r,p}$
by Theorem IV.4.1 of \cite{vishik1988mathematical}, so that the embedding of the former into the latter is compact. 

\vspace{15pt} 
Finally, to obtain tightness from Lemma II.3.1 of \cite{vishik1988mathematical}, we need the basic bound \eqref{tight-bd}, or 
\be \sup_{\nu>0}\int\left\lVert \bu\right\rVert_{\mathcal{E}^{\sigma,\beta,\gamma}_{r,p}}\,\mathrm{d}\mpr^\nu_\mu(\bu)<\infty. \lb{besov2} \ee 
Such a uniform estimate with the norm $\lVert\cdot\rVert_{L^r([0,T],B_{p,\infty}^\sigma(\mathbb{T}^d)}$ follows immediately from our main physical 
hypothesis \eqref{besov}. We spend the remainder of this subsection deriving the similar bound for the norm 
 $\lVert\cdot\rVert_{C^{0,\gamma}([0,T],B_{p/2,\infty}^{\sigma-\beta}(\mathbb{T}^d))}$ of the other component space in the definition of $\mathcal{E}^{\sigma,\beta,\gamma}_{r,p}.$ Toward this end, we define the quantity $\bz^\nu:=\bu^\nu-\nu^{\kappa}\mathscr{P}\grad\cdot \pi_\Lambda\bxx$,  in terms of which we can decompose the 
solutions of \eqref{llns} as 
$$ \bu^\nu = \bz^\nu + \nu^{\kappa}\mathscr{P}\grad\cdot \pi_\Lambda\bxx. $$
We obtain the required uniform estimate first for the term $\bz^\nu$ and then for the noise term $\nu^{\kappa}\mathscr{P}\grad\cdot \pi_\Lambda\bxx.$ 

\vspace{10pt} \noindent 
{\it Estimate for $\bz^\nu$:} It follows directly from the definition of $\bz^\nu$ that it satisfies the equation
\be \partial_t\bz^\nu=\mathscr{P}\pi_\Lambda\left[-\grad\cdot (\bu^\nu\otimes\bu^\nu)+\bsf\right]
+\nu\triangle \pi_\Lambda\bu^\nu. \lb{zeq} \ee 
We shall use the integral form of this equation 
\be \bz^\nu(\cdot,t)=\bz^\nu(\cdot,0)+\int_0^t\partial_t\bz^\nu(\cdot,s)\,\mathrm{d}s=\bu^\nu(\cdot,0)+\int_0^t\partial_t\bz^\nu(\cdot,s)\,\mathrm{d}s.\lb{zint} \ee 
to obtain our estimates. We proceed by a series of lemmas to control each term in \eqref{zeq}. 

\begin{lemma}\label{claim3} Take $2\leq s\leq\infty$, $2\leq p<\infty$, $0<\sigma<1$, $\beta\geq1$, and $\bvp\in L^s([0,T],B_{p,\infty}^{\sigma}(\mathbb{T}^d))$. Let
\[\forall t\in[0,T]:\hspace{.15in}\int_{\mathbb{T}^d}\bvp(\bx,t)\,\mathrm{d}\bx=0.\]
Then $\mathscr{P}\grad\cdot(\bvp\otimes\bvp)\in L^r([0,T],B_{p/2,\infty}^{\sigma-\beta}(\mathbb{T}^d))$ for any $1\leq r\leq s/2$ when $s<\infty$, and $\mathscr{P}\grad\cdot(\bvp\otimes\bvp)\in L^r([0,T],B_{p/2,\infty}^{\sigma-\beta}(\mathbb{T}^d))$ for any $1\leq r\leq\infty$ when $s=\infty$.\end{lemma}

\hfill

\textit{Proof.} To control the Leray projection $\mathscr{P},$ we use the fact that it is given by a matrix-valued Fourier multiplier $\mathscr{P}_{ij}(\bk)=
\delta_{ij}-k_ik_j/|\bk|^2$ which is homogeneous of degree 0. We employ the following specific estimate from Theorem 3.6.3 of \cite{schmeisser1987topics}: for $\mathscr{K}>d\left(\frac{1}{\min(p,1)}-\frac{1}{2}\right)$
\be \left\| \sum_{\bk\in {\mathbb Z}^d} m(\bk) ({\mathcal F}f)(\bk)\right\|_{B_{p,q}^{\sigma}(\mathbb{T}^d)}\leq c
\lVert m\rVert_{\mathfrak{h}^{\mathscr{K}}(\mathbb{R}^d)}\|f\|_{B_{p,q}^{\sigma}(\mathbb{T}^d)}, \lb{triebel} \ee 
where $c$ is a constant and Fourier multipliers $m$, considered as functions on $\mathbb{R}^d,$ are estimated by the norm 
\[\lVert m\rVert_{\mathfrak{h}^{\mathscr{K}}(\mathbb{R}^d)}:=\lVert\psi^\ast m\rVert_{H^{\mathscr{K}}(\mathbb{R}^d)}+\sup_{n\in\mathbb{N}}\,\lVert\psi(\cdot)m(2^n\cdot)\rVert_{H^{\mathscr{K}}(\mathbb{R}^d)}, \]
and where the smooth test functions $\psi^\ast,\psi\in\mathcal{S}(\mathbb{R}^d)$ with the properties 
\[\forall\bk:0\leq\psi^\ast(\bk)\leq1,\hspace{.15in}\operatorname{supp}\psi^\ast\subseteq\{\bk:|\bk|\leq4\},\hspace{.15in}\psi^\ast(\bk)=1\,\textrm{ when }\,|\bk|\leq2,\]\[\forall\bk:0\leq\psi(\bk)\leq1,\hspace{.15in}\operatorname{supp}\psi\subseteq\{\bk:1/4\leq|\bk|\leq4\},\hspace{.15in}\psi(\bk)=1\,\textrm{ when }\,1/2\leq|\bk|\leq2.\]
provide a spectral decomposition. 

In order to apply the above bound to the Leray projection, we must modify its multiplier to eliminate the singularity at $|\bk|=0,$ by taking 
\begin{equation}\label{lhproj}(\mathcal{F}\widetilde{\mathscr{P}}\bvp)_i(\mathbf{k}):=\sum_{j=1}^d \widetilde{\mathscr{P}}_{ij}(\bk)
(\mathcal{F}\bvp)_j(\mathbf{k}),  \quad \bk\in {\mathbb Z}^d \end{equation}
with modified matrix multiplier given by
\[ \widetilde{\mathscr{P}}_{ij}(\bk):=\begin{cases}\delta_{ij}-k_i k_j/|\bk|^2,\hspace{.15in}|\mathbf{k}|\geq1/4\\(\delta_{ij}-k_i k_j/|\mathbf{k}|^2)[1-\exp(-|\bk|^2/(1/16-|\mathbf{k}|^2))],\hspace{.15in}0<|\mathbf{k}|<1/4\\
0,\hspace{.3in}\mathbf{k}=0\end{cases}.\]
which is now $C^\infty$ and coincident with the original multiplier $\mathscr{P}_{ij}(\bk)$ for $|\bk|\geq 1/4.$ The only argument 
$\bk\in {\mathbb Z}^d$ in the expression \eqref{lhproj} affected by the modification is thus $\bk=\bzed,$ so that $\widetilde{\mathscr{P}}$
and $\mathscr{P}$ are identical in their action on fields $\bvp$ with zero space-mean. Hence, we can identify $\widetilde{\mathscr{P}}=\mathscr{P}$.
Now take $\mathscr{K}>d/2.$ Using that $\psi^\ast,\psi$ are smooth and vanish outside a compact set, and that $\widetilde{\mathscr{P}}_{ij}(2^n\bk)
=\widetilde{\mathscr{P}}_{ij}(\bk)$ for $|\bk|\geq1/4$, we conclude $\psi^\ast \widetilde{\mathscr{P}}_{ij},\psi(\cdot)\widetilde{\mathscr{P}}_{ij}(2^n\cdot)$ 
and their derivatives are uniformly bounded in $n$ and compactly supported. Thus $\lVert \widetilde{\mathscr{P}}\rVert_{\mathfrak{h}^{\mathscr{K}}(\mathbb{R}^d)}<\infty.$

By the estimate \eqref{triebel} we thus obtain with $\bv$ denoting $\bv(\cdot,t)$ for a.e. $t\in [0,T]$ that 
\[\lVert\mathscr{P}\grad\cdot(\bvp\otimes\bvp)\rVert_{B_{p/2,\infty}^{\sigma-\beta}(\mathbb{T}^d)}\lesssim\lVert\grad\cdot(\bvp\otimes\bvp)\rVert_{B_{p/2,\infty}^{\sigma-\beta}(\mathbb{T}^d)}.\]
Per Theorem 2.2 and Proposition 2.3 of \cite{sawano2018theory}, we have the estimate for $\beta\geq 1$
\[\lVert\grad\cdot(\bvp\otimes\bvp)\rVert_{B_{p/2,\infty}^{\sigma-\beta}(\mathbb{T}^d)}\lesssim\lVert\bvp\otimes\bvp\rVert_{B_{p/2,\infty}^{\sigma}(\mathbb{T}^d)}.\]
Since all finite-dimensional norms are equivalent, we use the natural inner product structure of tensored Hilbert space to write:
\[\left|\bvp\otimes\bvp\right|\lesssim\left|\langle\bvp\otimes\bvp,\bvp\otimes\bvp\rangle_{\mathbb{R}^3\otimes\mathbb{R}^3}\right|^{1/2}=\left|\langle\bvp,\bvp\rangle\right|=|\bvp|^2\hspace{0.15in}\implies\hspace{0.15in}\lVert\bvp\otimes\bvp\rVert_{L^{p/2}(\mathbb{T}^d)}\lesssim\lVert\bvp\rVert_{L^p(\mathbb{T}^d)}^2.\]
For any $i,j$ we have
\begin{align*}\lVert\bvp_i(\cdot+\mathbf{h})\bvp_j(\cdot+\mathbf{h})-\bvp_i\bvp_j\rVert_{L^{p/2}(\mathbb{T})^d}&\leq\lVert\bvp_i(\cdot+\mathbf{h})-\bvp_i\rVert_{L^p(\mathbb{T})^d}\lVert\bvp_j(\cdot+\mathbf{h})\rVert_{L^p(\mathbb{T})^d}\\
&\hspace{0.3in}+\lVert\bvp_i\rVert_{L^p(\mathbb{T})^d}\lVert\bvp_j(\cdot+\mathbf{h})-\bvp_j\rVert_{L^p(\mathbb{T})^d}.\end{align*}
All of the RHS terms are bounded by constants times $|\mathbf{h}|^\sigma,$ so that 
\[\lVert\bvp\otimes\bvp\rVert_{B_{p/2,\infty}^{\sigma}(\mathbb{T}^d)}\lesssim\lVert\bvp\rVert_{B_{p,\infty}^{\sigma}(\mathbb{T}^d)}^2.\]

For $s<\infty$, we use linearity and monotonicity of the Lebesgue integration over time, and then 
apply H\"older's inequality
\[\int_0^T\lVert\mathscr{P}\grad\cdot(\bvp\otimes\bvp)(\cdot,t)\rVert_{B_{p/2,\infty}^{\sigma-\beta}(\mathbb{T}^d)}^r\,\mathrm{d}t\lesssim\int_0^T\lVert\bvp(\cdot,t)\rVert_{B_{p,\infty}^{\sigma}(\mathbb{T}^d)}^{2r}\,\mathrm{d}t\lesssim\left(\int_0^T\lVert\bvp(\cdot,t)\rVert_{B_{p,\infty}^{\sigma}(\mathbb{T}^d)}^s\,\mathrm{d}t\right)^{2r/s},
\quad 1\leq r\leq s/2. \]
Thus
\[\left\lVert\mathscr{P}\grad\cdot(\bvp\otimes\bvp)\right\rVert_{L^r([0,T],B_{p/2,\infty}^{\sigma-\beta}(\mathbb{T}^d))}\lesssim\left\lVert\bvp\right\rVert_{L^s([0,T],B_{p,\infty}^{\sigma}(\mathbb{T}^d))}^2.\]
For $s=\infty$ we have
\[\underset{t\in[0,T]}{\operatorname{ess\,sup}}\,\left\lVert\mathscr{P}\grad\cdot\left[\bvp(\cdot,t)\otimes\bvp(\cdot,t)\right]\right\rVert_{B_{p/2,\infty}^{\sigma-\beta}(\mathbb{T}^d)}\lesssim\underset{t\in[0,T]}{\operatorname{ess\,sup}}\,\lVert\bvp(\cdot,t)\rVert^2_{B_{p,\infty}^{\sigma}(\mathbb{T}^d)},\]
which gives for $1\leq r\leq \infty$
\[ \left\lVert\mathscr{P}\grad\cdot(\bvp\otimes\bvp)\right\rVert_{L^r([0,T],B_{p/2,\infty}^{\sigma-\beta}(\mathbb{T}^d))}\lesssim 
\left\lVert\mathscr{P}\grad\cdot(\bvp\otimes\bvp)\right\rVert_{L^\infty([0,T],B_{p/2,\infty}^{\sigma-\beta}(\mathbb{T}^d))}\lesssim\left\lVert\bvp\right\rVert^2_{L^\infty([0,T],B_{p,\infty}^{\sigma}(\mathbb{T}^d))}. \qed \] 

\hfill

\begin{lemma}\label{claim4}Let $1\leq r\leq s\leq\infty$, $2\leq p<\infty$, $0<\sigma<1$, $\beta\geq2$, and $\bvp\in L^s([0,T],B_{p,\infty}^{\sigma}(\mathbb{T}^d))$. Then $\triangle\bvp\in L^r([0,T],B_{p/2,\infty}^{\sigma-\beta}(\mathbb{T}^d))$.\end{lemma}

\hfill

\textit{Proof.} Again from Theorem 2.2 and Proposition 2.3 of \cite{sawano2018theory}, we have 
\[\lVert\triangle\bvp(\cdot,t)\rVert_{B_{p/2,\infty}^{\sigma-\beta}(\mathbb{T}^d)}\lesssim\lVert\bvp(\cdot,t)\rVert_{B_{p/2,\infty}^{\sigma}(\mathbb{T}^d)}
\lesssim \lVert\bvp(\cdot,t)\rVert_{B_{p,\infty}^{\sigma}(\mathbb{T}^d)}, \]
using for the last step the continuous embedding $B_{p,\infty}^{\sigma}(\mathbb{T}^d)\hookrightarrow B_{p/2,\infty}^{\sigma}(\mathbb{T}^d)$. For $s<\infty$, we again apply the properties of the Lebesgue integral and H\"older inequality 
\[\int_0^T\lVert\triangle\bvp(\cdot,t)\rVert_{B_{p/2,\infty}^{\sigma-\beta}(\mathbb{T}^d)}^r\,\mathrm{d}t
\lesssim\int_0^T\lVert\bvp(\cdot,t)\rVert_{B_{p,\infty}^{\sigma}(\mathbb{T}^d)}^r\,\mathrm{d}t
\lesssim\left(\int_0^T\lVert\bvp(\cdot,t)\rVert_{B_{p,\infty}^{\sigma}(\mathbb{T}^d)}^s\,\mathrm{d}t\right)^{r/s}.\]
Thus
\[\left\lVert\triangle\bvp\right\rVert_{L^r([0,T],B_{p/2,\infty}^{\sigma-\beta}(\mathbb{T}^d))}\lesssim\left\lVert\bvp\right\rVert_{L^s([0,T],B_{p,\infty}^{\sigma}(\mathbb{T}^d))}.\]
Similarly for $s=\infty$ we have
\[\underset{t\in[0,T]}{\operatorname{ess\,sup}}\,\left\lVert\triangle\bvp(\cdot,t)\right\rVert_{B_{p/2,\infty}^{\sigma-\beta}(\mathbb{T}^d)}\lesssim\underset{t\in[0,T]}{\operatorname{ess\,sup}}\,\lVert\bvp(\cdot,t)\rVert_{B_{p,\infty}^{\sigma}(\mathbb{T}^d)},\]
and
\[\left\lVert\triangle\bvp\right\rVert_{L^r([0,T],B_{p/2,\infty}^{\sigma-\beta}(\mathbb{T}^d))}\lesssim
\left\lVert\triangle\bvp\right\rVert_{L^\infty([0,T],B_{p/2,\infty}^{\sigma-\beta}(\mathbb{T}^d))}\lesssim
\left\lVert\bvp\right\rVert_{L^\infty([0,T],B_{p,\infty}^{\sigma}(\mathbb{T}^d))}, \quad 1\leq r\leq\infty \qed \]

\hfill

Finally, to control the effect of the projections $\pi_\Lambda,$ we have 

\hfill

\begin{lemma}\label{claim5}For any $\sigma\in {\mathbb R}$, $1\leq p\leq\infty$, 
$\bv\in B^\sigma_{p,\infty}({\mathbb T}^d)$ and $\Lambda=2^m$ for some $m\in\mathbb{N}$, we have
\[\lVert \pi_\Lambda\bvp\rVert_{B_{p,\infty}^{\sigma}(\mathbb{T}^d)}\uparrow \lVert\bvp\rVert_{B_{p,\infty}^{\sigma}(\mathbb{T}^d)}
\quad \mbox { as } \quad  m\uparrow\infty. 
\]
\end{lemma}

\hfill

\textit{Proof.} Define corridors in Fourier space as 
\[C_0:=\{\mathbf{k}:|\mathbf{k}|_\infty<1\},\hspace{.15in}C_j:=\{\mathbf{k}:2^{j-1}\leq|\mathbf{k}|_\infty<2^j\}.\]
It then suffices to recall the Lizorkin representations (Theorem 3.5.3 of \cite{schmeisser1987topics}):
\[\lVert\bvp\rVert_{B_{p,\infty}^{\sigma}(\mathbb{T}^d)}=\sup_{j\in\mathbb{N}}2^{\sigma j}\left\lVert\sum_{\mathbf{k}\in C_j}(\mathcal{F}\bvp)(\mathbf{k})\exp\left(i\langle \mathbf{k},\cdot\rangle\right)\right\rVert_{L^p(\mathbb{T}^d)}.\]
Since $(\mathcal{F}\bvp)(\mathbf{k})$ is just the coefficient of the $\mathbf{k}$th eigenbasis vector, when $|\mathbf{k}|_\infty<2^m$ and $|\mathbf{k}|_\infty\geq 2^m$ we have $(\mathcal{F} \pi_\Lambda\bvp)(\mathbf{k})=(\mathcal{F}\bvp)(\mathbf{k})$ and $(\mathcal{F} \pi_\Lambda\bvp)(\mathbf{k})=0$ respectively, and the sum in the Lizorkin representation vanishes on $C_j$ when $j>m$. Thus
\[\lVert \pi_\Lambda\bvp\rVert_{B_{p,\infty}^{\sigma}(\mathbb{T}^d)}=\sup_{j\leq m}2^{\sigma j}\left\lVert\sum_{\mathbf{k}\in C_j}(\mathcal{F}\bvp)(\bk)\exp\left(i\langle\mathbf{k},\cdot\rangle\right)\right\rVert_{L^p(\mathbb{T}^d)}\]
and monotonic convergence of $\lVert \pi_\Lambda\bvp\rVert_{B_{p,\infty}^{\sigma}(\mathbb{T}^d)}$ to $\lVert\bvp\rVert_{B_{p,\infty}^{\sigma}(\mathbb{T}^d)}$
becomes obvious. $\qed$

\hfill

Combining all of these results, we infer from Lemma~\ref{claim3} and Lemma~\ref{claim5} that
\[\lVert\mathscr{P}\grad\cdot \pi_\Lambda(\bu^\nu\otimes\bu^\nu)\rVert_{L^{r/2}([0,T],B_{p/2,\infty}^{\sigma-\beta}(\mathbb{T}^d))}\lesssim\lVert\bu^\nu\rVert_{L^r([0,T],B_{p,\infty}^{\sigma}(\mathbb{T}^d))}^2,\]
and from Lemma~\ref{claim4} and Lemma~\ref{claim5} that
\[\lVert\triangle \pi_\Lambda\bu^\nu\rVert_{L^{r/2}([0,T],B_{p/2,\infty}^{\sigma-\beta}(\mathbb{T}^d))}\lesssim\lVert\bu^\nu\rVert_{L^r([0,T],B_{p,\infty}^{\sigma}(\mathbb{T}^d))}.\]
Finally, by Lemma~\ref{claim5} we evidently have
\[\lVert \pi_\Lambda\bsf\rVert_{L^{r/2}([0,T],B_{p/2,\infty}^{\sigma-\beta}(\mathbb{T}^d))}\lesssim\lVert\bsf\rVert_{C^0([0,T],B_{p/2,\infty}^{\sigma-\beta}(\mathbb{T}^d))}.\]
As all of our estimates above are ${\mathbb P}\times \mu$ almost sure, we get 
$$ \|\partial_t\bz^\nu\|_{L^{r/2}([0,T],B_{p/2,\infty}^{\sigma-\beta}(\mathbb{T}^d))}  
\lesssim 
\lVert \bu^\nu\rVert_{L^r([0,T],B_{p,\infty}^{\sigma}(\mathbb{T}^d))}^2+\lVert \bu^\nu\rVert_{L^r([0,T],B_{p,\infty}^{\sigma}(\mathbb{T}^d))}
+\lVert \bsf\rVert_{C^0([0,T],B_{p/2,\infty}^{\sigma-\beta}(\mathbb{T}^d))}
$$
almost surely. 

Next we define $\Delta_{t,s}\bz^\nu :=\bz^\nu(\cdot,t)-\bz^\nu(\cdot,s)$ and note by \eqref{zint} and H\"older inequality that 
\begin{align*}\lVert \Delta_{t,s}\bz^\nu \rVert_{B_{p/2,\infty}^{\sigma-\beta}(\mathbb{T}^d)}&
\leq\int_s^t\lVert\partial_t\bz^\nu(\cdot,\tau) \rVert_{B_{p/2,\infty}^{\sigma-\beta}(\mathbb{T}^d)}\,\mathrm{d}\tau\\
&\lesssim|t-s|^{(r-2)/r}\,\lVert\partial_t\bz^\nu\rVert_{L^{r/2}([0,T],B_{p/2,\infty}^{\sigma-\beta}(\mathbb{T}^d))}.\end{align*}
Recalling the definition of the semi-norm 
$$ |\!|\!| \bz |\!|\!|_{C^{0,\gamma}([0,T], B_{p/2,\infty}^{\sigma-\beta}(\mathbb{T}^d))}:=
\sup_{t,s\in[0,T]}\frac{\lVert \bz(\cdot,t)-\bz(\cdot,s)\rVert_{B_{p/2,\infty}^{\sigma-\beta}(\mathbb{T}^d)}}{|t-s|^{\gamma}} $$
we obtain for $\gamma\leq (r-2)/r$ 
\begin{align*}{\mathbb E}_\mu\left[ |\!|\!| \bz^\nu |\!|\!|_{C^{0,\gamma}([0,T], B_{p/2,\infty}^{\sigma-\beta}(\mathbb{T}^d))} \right]  
&\lesssim {\mathbb E}_\mu\left[ \lVert\partial_t\bz^\nu\rVert_{L^{r/2}([0,T],B_{p/2,\infty}^{\sigma-\beta}(\mathbb{T}^d))}\right] \\
&\lesssim\int\lVert \bu\rVert_{L^r([0,T],B_{p,\infty}^{\sigma}(\mathbb{T}^d))}^2\,\mathrm{d}\mpr^\nu_\mu(\bu)+\left(\int\lVert \bu\rVert_{L^r([0,T],B_{p,\infty}^{\sigma}(\mathbb{T}^d))}^2\,\mathrm{d}\mpr^\nu_\mu(\bu)\right)^{1/2}\\
&\hspace{.5in}+\lVert \bsf\rVert_{C^0([0,T],B_{p/2,\infty}^{\sigma-\beta}(\mathbb{T}^d))},\end{align*}

We need a similar bound for the $C^{0}([0,T], B_{p/2,\infty}^{\sigma-\beta}(\mathbb{T}^d))$ norm. We use the expression 
\eqref{zint} and bound separately the contributions from $\bz^\nu(\cdot,0)$ and $\int_0^t\partial_t\bz^\nu(\cdot,s)\,\mathrm{d}s.$ From Lemmas \ref{claim2} and \ref{claim5} and 
from our assumption $\bu_0\in L^p(\mathbb{T}^d),$
we have $\bu^\nu(\cdot,0)=\pi_\Lambda\bu_0$ bounded uniformly across $\nu$ in $B_{p/2,\infty}^{\sigma-\beta}(\mathbb{T}^d).$ Then we have
\[\left\lVert\int_0^t\partial_t\bz^\nu(\cdot,s)\,\mathrm{d}s\right\rVert_{C^0([0,T],B_{p/2,\infty}^{\sigma-\beta}(\mathbb{T}^d))}\leq\int_0^T\left\lVert\partial_t\bz^\nu(\cdot,s)\right\rVert_{B_{p/2,\infty}^{\sigma-\beta}(\mathbb{T}^d)}\,\mathrm{d}s\lesssim\lVert\partial_t\bz^\nu\rVert_{L^{r/2}([0,T],B_{p/2,\infty}^{\sigma-\beta}(\mathbb{T}^d))}.\]
We have already estimated the last term, and all the estimates are almost sure in ${\mathbb P}\times \mu$. Thus we obtain 
\begin{align*}
{\mathbb E}_\mu\left[\lVert\bz^\nu\rVert_{C^0([0,T],B_{p/2,\infty}^{\sigma-\beta}(\mathbb{T}^d))}\right]&\lesssim\int\lVert \bu\rVert_{L^r([0,T],B_{p,\infty}^{\sigma}(\mathbb{T}^d))}^2\,\mathrm{d}\mpr^\nu_\mu(\bu)+\left(\int\lVert \bu\rVert_{L^r([0,T],B_{p,\infty}^{\sigma}(\mathbb{T}^d))}^2\,\mathrm{d}\mpr^\nu_\mu(\bu)\right)^{1/2}\\
&\hspace{.5in}+\lVert \bsf\rVert_{C^0([0,T],B_{p/2,\infty}^{\sigma-\beta}(\mathbb{T}^d))}+\lVert\bu_0\rVert_{L^p(\mathbb{T}^d)}.\end{align*}
Finally, we conclude from our empirical hypothesis \eqref{besov} that for $\gamma\leq (r-2)/r$ 
$$ \sup_{\nu>0} {\mathbb E}_\mu\left[\lVert\bz^\nu\rVert_{C^{0,\gamma} ([0,T],B_{p/2,\infty}^{\sigma-\beta}(\mathbb{T}^d))}\right]<\infty. $$

\hfill  

\vspace{10pt}\noindent 
{\it Estimate of $\mathscr{P}\grad\cdot \pi_\Lambda\bxx$:}
Next, we argue as before from Lemma~\ref{claim3} that
\[\left\lVert\mathscr{P}\grad\cdot[ \pi_\Lambda\Delta_{t,s}\bxx]\right\rVert_{B_{p/2,\infty}^{\sigma-\beta}(\mathbb{T}^d)}
\lesssim\left\lVert \pi_\Lambda\Delta_{t,s}\bxx\right\rVert_{B_{p/2,\infty}^{\sigma-\beta+1}(\mathbb{T}^d)}.\]
Using the representation \eqref{xidef}, we can write  
\[ \pi_\Lambda\Delta_{t,s}\bxx^\bk = \sum_{|\bk|_\infty\leq2^{m(\nu)}}\Delta_{t,s}\bxx^\bk\exp(i\langle \bk,\cdot\rangle ,\]
where the complex matrix-valued Brownian motions $\bxx^\bk$ are mutually independent except for the constraints $\overline{\xi_{ij}^\bk(t)}
=\xi_{ij}^{-\bk}(t),$ $\xi_{ji}^\bk(t)=\xi_{ij}^\bk(t)$ and $\sum_{i=1}^d \xi_{ii}^\bk(t)=0.$ Using a Littlewood--Paley form of the Besov norm,  
\[\lVert \pi_\Lambda\Delta_{t,s}\bxx^\bk\rVert_{B_{p/2,\infty}^{\sigma-\beta+1}(\mathbb{T}^d)} = \sup_{j\in\mathbb{N}} \ 2^{(\sigma-\beta+1)j}  \left\lVert\sum_{|\bk|_\infty\leq  2^{m(\nu)}}\varphi_j(\bk)\Delta_{t,s}\bxx^\bk\exp(i\langle \bk,\cdot\rangle_{\mathbb{R}^d})\right\rVert_{L^{p/2}(\mathbb{T}^d)},\]
where $(\varphi_j|j\in {\mathbb N})$ is a smooth partition of unity such that the $\varphi_j\in \mathcal{S}({\mathbb R}^d)$ are uniformly bounded 
and supported on $2^{j-1}<|\bk|_\infty<2^{j+1}.$ See \cite{schmeisser1987topics}, section 3.5.1. From the Minkowski inequality,
\[\left\lVert\sum_{|\bk|_\infty\leq 2^{m(\nu)}}\varphi_j(\bk)\Delta_{t,s}\bxx^\bk\exp(i\langle \bk,\cdot\rangle_{\mathbb{R}^d})\right\rVert_{L^{p/2}(\mathbb{T}^d)}\leq\sum_{|\bk|_\infty\leq 2^{m(\nu)}}\varphi_j(\bk)|\Delta_{t,s}\bxx^\bk|\lVert\exp(i\langle \bk,\cdot\rangle_{\mathbb{R}^d})\rVert_{L^{p/2}(\mathbb{T}^d)},\]
where $\lVert\exp(i\langle \bk,\cdot\rangle_{\mathbb{R}^d})\rVert_{L^{p/2}(\mathbb{T}^d)}$ is constant in $\bk$. Thus
\[\lVert \pi_\Lambda \Delta_{t,s}\bxx^\bk\rVert_{B_{p/2}^{\sigma-\beta+1,\infty}(\mathbb{T}^d)}\lesssim\sup_{j\in\mathbb{N}}2^{(\sigma-\beta+1)j}\sum_{|\bk|_\infty\leq 2^{m(\nu)}}\varphi_j(\bk)|\Delta_{t,s}\bxx^\bk|,\]
and we get
\begin{align*}\sup_{j\in\mathbb{N}}2^{(\sigma-\beta+1)j}\sum_{|k|_\infty\leq 2^{m(\nu)}}\varphi_j(\bk)|\Delta_{t,s}\bxx^\bk|&\lesssim\,\sup_{j\in\mathbb{N}}2^{(\sigma-\beta+1)j}\sum_{|\bk|_\infty\leq 2^{m(\nu)}}\boldsymbol{1}_{\{2^{j-1}<|\bk|_\infty<2^{j+1}\}}(\bk)|\Delta_{t,s}\bxx^\bk|\\
&\lesssim\sup_{j\leq m(\nu)}2^{(\sigma-\beta+1+d)j}\max_{2^{j-1}<|\bk|_\infty\leq2^{j+1}}|\Delta_{t,s}\bxx^\bk|.\end{align*}
From the definition of the H\"older semi-norm $|\! |\!| f  |\! |\!|_{C^{0,\gamma}([0,T],B_{p/2,\infty}^{\sigma-\beta}(\mathbb{T}^d))}:=
\sup_{t\neq s\in [0,T]} \frac{\left\lVert \Delta_{t,s}f\right\rVert_{B_{p/2,\infty}^{\sigma-\beta}(\mathbb{T}^d)}}{|t-s|^\gamma},$ we thus obtain finally 
\bea  |\! |\!| \mathscr{P}\grad\cdot \pi_\Lambda\bxx  |\! |\!|_{C^{0,\gamma}([0,T],B_{p/2,\infty}^{\sigma-\beta}(\mathbb{T}^d))}
&\lesssim & \sup_{j<m(\nu)} 2^{(\sigma-\beta+1+d)j} \max_{2^{j-1}<|\bk|_\infty\leq2^{j+1}}\lVert \bxx^\bk\rVert_{C^{0,\gamma}([0,T],\mathbb{R}^d)} \cr 
&\leq & \sum_{j=0}^{m(\nu)}2^{(\sigma-\beta+1+d)j} \max_{2^{j-1}<|\bk|_\infty\leq2^{j+1}}\lVert \bxx^\bk\rVert_{C^{0,\gamma}([0,T],\mathbb{R}^d)} \eea 
Note that all of our estimates for $\mathscr{P}\grad\cdot \pi_\Lambda \bxx$ are ${\mathbb P}$-almost sure.

Taking expectations
\[{\mathbb E}\left[|\! |\!| \mathscr{P}\grad\cdot \pi_\Lambda\bxx  |\! |\!|_{C^{0,\gamma}([0,T],B_{p/2,\infty}^{\sigma-\beta}(\mathbb{T}^d))}\right]\leq\sum_{j=0}^{m(\nu)}2^{(\sigma-\beta+1+d)j}{\mathbb E}\left[\max_{2^{j-1}<|\bk|_\infty\leq2^{j+1}}\lVert \bxx^\bk\rVert_{C^{0,\gamma}([0,T],\mathbb{R}^d)}\right].\]
However, 
\begin{align*}{\mathbb E}\left[\max_{2^{j-1}<|\bk|_\infty\leq2^{j+1}}\lVert \bxx^\bk\rVert_{C^{0,\gamma}([0,T],\mathbb{R}^d)}\right]&\leq\sum_{2^{j-1}<|\bk|_\infty\leq2^{j+1}}{\mathbb E}\left[\lVert \bxx^\bk\rVert_{C^{0,\gamma}([0,T],\mathbb{R}^d)}\right]\\
&\lesssim \hspace{20pt} 2^{jd}\ {\mathbb E}\left[\lVert \widetilde{\bxx}\rVert_{C^{0,\gamma}([0,T],\mathbb{R}^d)}\right].\end{align*}
where in the last line we used the fact that the random variables $\bxx^\bk$ for $\bk\neq\bzed$ are all identically distributed and 
$\widetilde{\bxx}$ is a single representative.
We now recall the classical result that ${\mathbb E}\left[\lVert \widetilde{\bxx}\rVert^n_{C^{0,\gamma}([0,T],\mathbb{R}^d)}\right]<\infty$ for any $\gamma<1/2$ and $n\in {\mathbb N}$
and, in fact, ${\mathbb E}\left[e^{\epsilon \lVert \widetilde{\bxx}\rVert_{C^{0,\gamma}([0,T],\mathbb{R}^d)}}\right]<\infty$ for sufficiently small $\epsilon>0$ \cite{veraar2008besov}. The latter statement is essentially a consequence of Fernique's theorem \cite{fernique1970integrabilite}, modulo some 
technical issues due to the non-separability of the H\"older space $C^{0,\gamma}([0,T],\mathbb{R}^d);$ e.g. see \cite{baldi1992large}.  From  
this fact and the condition $\sigma-\beta+1+2d<0,$ we thus deduce for any $\gamma<1/2$ that 
\begin{align*}{\mathbb E}\left[|\! |\!| \mathscr{P}\grad\cdot \pi_\Lambda\bxx  |\! |\!|_{C^{0,\gamma}([0,T],B_{p/2,\infty}^{\sigma-\beta}(\mathbb{T}^d))}
\right]&\lesssim\sum_{j=0}^{m(\nu)}2^{(\sigma-\beta+1+2d)j}\,\,\propto\,\,1-\nu^{-(\sigma-\beta+1+2d)\alpha}.\end{align*}
so that the lefthand side is uniformly bounded in $\nu$. 

To get a similar estimate for the $C^{0}([0,T],\mathbb{R}^d)$ norm, we use an almost identical argument, which yields  
\[{\mathbb E}\left[\lVert \mathscr{P}\grad\cdot \pi_\Lambda\bxx  \rVert_{C^{0}([0,T],B_{p/2,\infty}^{\sigma-\beta}(\mathbb{T}^d))}\right]\leq\sum_{j=0}^{m(\nu)}2^{(\sigma-\beta+1+2d)j} \ {\mathbb E}\left[\lVert \widetilde{\bxx}\rVert_{C^{0}([0,T],\mathbb{R}^d)}\right].\]
The boundedness of the latter expectation is an even more classical result in probability theory, which can be deduced from Doob's maximal inequality 
for submartingales or alternatively from more elementary considerations for Brownian motion. We obtain once again a bound uniform in $\nu.$

Finally, we can conclude that for any $\gamma<1/2$
\[\sup_{\nu>0}{\mathbb E}\left[\left\lVert\mathscr{P}\grad\cdot \pi_\Lambda \bxx\right\rVert_{C^{0,\gamma}([0,T],B_{p/2,\infty}^{\sigma-\beta}(\mathbb{T}^d))}\right]<\infty.\]
It is noteworthy that the prefactor $\nu^{\kappa}$ played no role in this estimate. 

\hfill   

\vspace{10pt}\noindent 
{\it Estimate of $\bu^\nu= \bz^\nu + \nu^{\kappa}\mathscr{P}\grad\cdot \pi_\Lambda\bxx$:} Putting together the previous bounds, we 
obtain
\[\sup_{\nu>0}\int\left\lVert\bu\right\rVert_{C^{0,\gamma}([0,T],B_{p/2,\infty}^{\sigma-\beta}(\mathbb{T}^d))}\,\mathrm{d}\mpr^\nu_\mu(\bu)
=\sup_{\nu>0}{\mathbb E}_\mu \left[\left\lVert\bu^\nu\right\rVert_{C^{0,\gamma}([0,T],B_{p/2,\infty}^{\sigma-\beta}(\mathbb{T}^d))}\right]
<\infty,\]
for any $\gamma<\min\{(r-2)/r,1/2\}.$ Combined with our hypothesis \eqref{besov}, we then get that
\[\sup_{\nu>0}\int\left\lVert \bu\right\rVert_{\mathcal{E}^{\sigma,\beta,\gamma}_{r,p}}\,\mathrm{d}\mpr^\nu_\mu(\bu)<\infty.\]
By Lemma II.3.1 of \cite{vishik1988mathematical}, we conclude that the family of measures $\mpr^\nu_\mu,$ $\nu>0$ are tight  
on $\mathcal{Z}^{\sigma,\beta}_{r,p}$ and there exists a subsequence $\nu_k\to0$ so that 
$\mpr^{\nu_k}_\mu\to \mpr_\mu$ weakly.

\hfill

\subsection{Proof of (ii)} We establish here support properties of the weak limit measures in (i). In particular, we will show 
that $P_\mu$ is supported on a subset of $\mathcal{Z}^{\sigma,\beta}_{r,p},$ so that realizations have some spatial Besov regularity.  
Support on $C^0([0,T],B_{p/2,\infty}^{\sigma-\beta}(\mathbb{T}^d))$ is immediate from (i), so that we focus on showing that $P_\mu$ is supported 
on the measurable subset $L^r([0,T],B_{p,\infty}^{\sigma}(\mathbb{T}^d))$ of $L^r([0,T],L^p(\mathbb{T}^d)).$ Note that measurability 
follows from Lemma~\ref{claim1}. We proceed analogously to Theorem IV.5.1 of \cite{vishik1988mathematical}, making 
use of their simple lemma: 

\vspace{10pt} 
\noindent 
{\bf Lemma II.3.2 of \cite{vishik1988mathematical}}: {\it Let $\Omega$ be a metric space and measures $\mu,$ $\mu_n,$ 
$n\in {\mathbb N}$ be defined on  ${\mathcal B}(\Omega)$ and $\mu_n\to\mu$ weakly as $n\to\infty.$ Let $f(u)$ be a 
continuous function on $\Omega$ and $\sup_n \int |f(u)| d\mu_n(u) \leq C.$ Then, $\int |f(u)| d\mu(u) \leq C.$}
\vspace{10pt} 

We first claim that $\bu\mapsto\lVert \pi_\Lambda \bu\rVert_{L^r([0,T],B_{p,\infty}^{\sigma}(\mathbb{T}^d))}$ for $\Lambda\equiv 2^m$
and $m\in{\mathbb N}$ is continuous on $\mathcal{Z}^{\sigma,\beta}_{r,p}.$ 
It suffices to observe the following bound by way of the Lizorkin representation:
\begin{align*}\lVert \pi_\Lambda \bu\rVert_{B_{p,\infty}^{\sigma}(\mathbb{T}^d)}&=\sup_{j<m(\nu)}2^{\sigma j}\left\lVert\sum_{k\in C_j}(\mathcal{F}\bu)(k)\exp(i\langle \bk,\cdot\rangle)\right\rVert_{L^p(\mathbb{T}^d)}\\
&\leq 2^{\sigma m(\nu)}\left\lVert\left[\sum_{j<m(\nu)}\left|\sum_{k\in C_j}(\mathcal{F}\bu)(\bk)\exp(i\langle \bk,\cdot\rangle)\right|^2\right]^{1/2}\right\rVert_{L^p(\mathbb{T}^d)}\\
&\leq 2^{\sigma m(\nu)}\lVert \bu\rVert_{F^{0}_{p,2}(\mathbb{T}^d)}\ \lesssim\lVert \bu\rVert_{L^p(\mathbb{T}^d)}.\end{align*}
The last estimate follows by the Paley--Littlewood theorem (Remark 2 of 3.5.4 of \cite{schmeisser1987topics}). 

Now by Lemma~\ref{claim5} the map $m \mapsto\lVert \pi_{2^m}\bu\rVert_{L^r([0,T],B_{p,\infty}^{\sigma}(\mathbb{T}^d))}$ is nondecreasing.
Fixing $m\in\mathbb{N}^+$ and setting $\Lambda\equiv2^m$, we have:
\[\int\lVert \pi_\Lambda \bu\rVert_{L^r([0,T],B_{p,\infty}^{\sigma}(\mathbb{T}^d))}\,\mathrm{d}\mpr^\nu_\mu(\bu)\leq\sup_{\nu>0}\int\lVert \bu\rVert_{L^r([0,T],B_{p,\infty}^{\sigma}(\mathbb{T}^d))}\,\mathrm{d}\mpr^\nu_\mu(\bu):=M.\]
Then by Lemma II.3.2 of \cite{vishik1988mathematical} we have for any weak limit measure 
\[\int\lVert \pi_\Lambda \bu\rVert_{L^r([0,T],B_{p,\infty}^{\sigma}(\mathbb{T}^d))}\,\mathrm{d}\mpr_\mu(\bu)\leq M.\]
Finally, a Beppo-Levi monotone convergence argument gives us
\[\int\lVert \bu\rVert_{L^r([0,T],B_{p,\infty}^{\sigma}(\mathbb{T}^d))}\,\mathrm{d}\mpr_\mu(\bu)
=\lim_{\Lambda\uparrow\infty}\int\lVert \pi_\Lambda \bu\rVert_{L^r([0,T],B_{p,\infty}^{\sigma}(\mathbb{T}^d))}\,\mathrm{d}\mpr_\mu(\bu)\leq M.\]
We conclude that $\mpr_\mu$ is supported on $\widetilde{\mathcal{E}}^{\sigma,\beta}_{r,p}:=L^r([0,T],B_{p,\infty}^{\sigma}(\mathbb{T}^d))\cap C^0([0,T],B_{p/2,\infty}^{\sigma-\beta}(\mathbb{T}^d))$. 

\hfill

We next show that $P_\mu$ is supported on zero-mean functions, i.e. 
\[ P_\mu\left(\left\{\bu:\bu\in\mathcal{Z}^{\sigma,\beta}_{r,p}\textrm{ and 
}\int_{\mathbb{T}^d}\bu(\bx,t)\,\mathrm{d}^dx=\mathbf{0}\textrm{ for every }t\in[0,T]\right\}\right)=1.\]
The map
\[\bu\mapsto\min\left(\underset{t\in[0,T]}{\operatorname{max}}\left|\int_{\mathbb{T}^d}\bu(\bx,t)\,\mathrm{d}^d x\right|,1\right)\]
is continuous on $\mathcal{Z}^{\sigma,\beta}_{r,p}\subseteq C^0([0,T],B_{p/2,\infty}^{\sigma-\beta}(\mathbb{T}^d))$ 
since $1\in B_{(p-2)/p,1}^{\beta-\sigma}(\mathbb{T}^d)),$ the Banach dual of $B_{p/2,\infty}^{\sigma-\beta}(\mathbb{T}^d))$ 
[Section 3.5.6 of \cite{schmeisser1987topics}]. The map is also evidently bounded. Thus we apply part (i) and observe taking $\nu\to 0$
\[0=\int\min\left(\underset{t\in[0,T]}{\operatorname{max}}\left|\int_{\mathbb{T}^d}\bu(\bx,t)\,\mathrm{d}^d x\right|,1\right)\,\mathrm{d}\mpr^\nu_\mu(\bu)\to\int\min\left(\underset{t\in[0,T]}{\operatorname{max}}\left|\int_{\mathbb{T}^d}\bu(\bx,t)\,\mathrm{d}^d x\right|,1\right)\,\mathrm{d}\mpr_\mu(\bu)=0.\]
The integrand on the RHS is nonnegative so that we conclude that for $\mpr_\mu-$a.e. $\bu$
\[\min\left(\underset{t\in[0,T]}{\operatorname{max}}\left|\int_{\mathbb{T}^d}\bu(\bx,t)\,\mathrm{d}^d x\right|,1\right)=0\hspace{.3in}\implies\hspace{.3in}\forall t\in[0,T]:\hspace{.15in}\int_{\mathbb{T}^d}\bu(\bx,t)\,\mathrm{d}^d x=\mathbf{0}.\]

\subsection{Proof of (iii)} We prove here that realizations of the limiting measure are weak solutions 
of the incompressible Euler equations. We follow the basic strategy of \cite{vishik1988mathematical},
who constructed space-time statistical solutions of incompressible Navier-Stokes, using in their 
Theorem IV.5.2 the idea of a ``Leray residual''. 
Similarly, we define here for $t\in [0,T],$ $\bsf\in C^0([0,T]\times \mathbb{T}^d),$ 
the {\it Euler residual} $E_t^\bsf:\mathcal{Z}^{\sigma,\beta}_{r,p}\times
{\mathcal S}(\mathbb{T}^d)\to\mathbb{R}$ 
\begin{equation}\label{bieu}E_t^\bsf(\bu,\bph):=\left\langle \bu(\cdot,t),\bph\right\rangle-\left\langle \bu(\cdot,0),\bph\right\rangle+\int_0^t\left\langle\mathscr{P}\grad\cdot(\bu\otimes \bu)(\cdot,s),\bph\right\rangle\,\mathrm{d}s-\int_0^t\left\langle \mathscr{P}\bsf(\cdot,s),\bph\right\rangle\,\mathrm{d}s.\end{equation}
The conditions
\be E_t^\bsf(\bu,\bph)=0, \quad \forall t\in [0,T], \ \forall \bph\in {\mathcal S}({\mathbb T}^d) \lb{mild} \ee 
thus state that the velocity field $\bu$ is a solution of Euler equations in a suitable weak sense, sometimes termed a ``mild solution.'' 
Note that above and hereafter we use ${\mathcal S}({\mathbb T}^d)$ as a simplified notation for ${\mathcal S}({\mathbb T}^d,{\mathbb R}^d).$ 
The dual product in the first two terms of \eqref{bieu} may be taken as an element of $B_{p/2,\infty}^{\sigma-\beta}(\mathbb{T}^d)=(B_{2/(2-p),1}^{\beta-\sigma}(\mathbb{T}^d))^*$ 
acting on an element of its predual, since $\mathcal{S}(\mathbb{T}^d)\subseteq B_{2/(2-p),1}^{\beta-\sigma}(\mathbb{T}^d).$
The same is true in the fourth term, where $\left\langle \bsf(\cdot,s),\mathscr{P}\bph\right\rangle$ is then continuous in $s,$
so that the integral is well-defined. Finally, for the third term, we note from (\ref{lhproj}) that $({\mathcal F}\widetilde{{\mathscr P}}\bph)(\bk)$ decays faster than any inverse power of $|\bk|$ if $\bph\in\mathcal{S}(\mathbb{T}^d)$. Hence, $\widetilde{\mathscr{P}}\bph\in\mathcal{S}(\mathbb{T}^d)$ and in particular $\grad\mathscr{P}\bph$ is a bounded function on $\mathbb{T}^d$. Thus, this term is well-defined for any element $\bu\in L^r([0,T],L^p(\mathbb{T}^d))\subset 
\mathcal{Z}^{\sigma,\beta}_{r,p}.$ 

\hfill 

We note some basic properties of the Euler residual that are needed below: 
\begin{lemma}\label{claim6} 
The following continuity and boundedness properties hold for all $\bsf\in C^0([0,T]\times \mathbb{T}^d)$: \\ 
(i) For all fixed $t\in [0,T],$ $\bph\in {\mathcal S}(\mathbb{T}^d),$ the map $\bu\mapsto E_t^\bsf(\bu,\bph)$ is strongly
continuous on $\mathcal{Z}^{\sigma,\beta}_{r,p}$.\\
(ii)  For all fixed $t\in [0,T],$ $\bu\in \mathcal{Z}^{\sigma,\beta}_{r,p},$ the map $\bph\mapsto E_t^\bsf(\bu,\bph)$ is linear and 
continuous on $\mathcal{S}(\mathbb{T}^d);$\\
(iii) For all fixed $\bu\in \mathcal{Z}^{\sigma,\beta}_{r,p},$ $\bph\in {\mathcal S}(\mathbb{T}^d),$ the map $t \mapsto E_t^\bsf(\bu,\bph)$
is continuous on $[0,T].$\\
(iv) For all fixed $t\in [0,T],$  $\bph\in\mathcal{S}(\mathbb{T}^d)$ and for some positive constants $c,$ $c',$ $c''$ dependent on those choices
$$ |E_t^\bsf(\bu,\bph)|\leq c +c' \|\bu\|_{\mathcal{Z}^{\sigma,\beta}_{r,p}} +c'' \|\bu\|_{\mathcal{Z}^{\sigma,\beta}_{r,p}}^2 $$
\end{lemma}

\hfill

\textit{Proof.} (i) Consider a sequence $\bu_n\to \bu$ strong in $\mathcal{Z}^{\sigma,\beta}_{r,p}.$ 
Noting $\mathcal{Z}^{\sigma,\beta}_{r,p}\subseteq C^0([0,T],B_{p/2,\infty}^{\sigma-\beta}(\mathbb{T}^d)),$ $\langle \bu_n(\cdot,s),\bph\rangle\to\langle \bu(\cdot,s),\bph\rangle$ for any $s\in[0,T]$ and $\bph\in \mathcal{S}(\mathbb{T}^d)\subseteq B_{2/(2-p),1}^{\beta-\sigma}(\mathbb{T}^d),$ since strong convergence in $B_{p/2,\infty}^{\sigma-\beta}(\mathbb{T}^d))$ implies weak-$\ast$ convergence and uniform convergence in $[0,T]$ implies pointwise convergence. 
For the third term, we observe the inequalities 
\begin{align*}
\left|\int_0^t\left\langle\mathscr{P}\grad\cdot(\bu_n\otimes \bu_n-\bu\otimes \bu)(\cdot,s),\bph\right\rangle\,\mathrm{d}s\right|&
=\left|\int_0^t\left\langle \bu_n\otimes \bu_n-\bu\otimes \bu,\grad\mathscr{P}\bph\right\rangle\,\mathrm{d}s\right|\\
&=\left|\int_0^t\left\langle \bu_n\otimes(\bu_n-\bu)-(\bu-\bu_n)\otimes \bu,\grad\mathscr{P}\bph\right\rangle\,\mathrm{d}s\right|\\
&\lesssim\int_0^t\left\lVert \bu_n\right\rVert_{L^2(\mathbb{T}^d)}\left\lVert \bu_n-\bu\right\rVert_{L^2(\mathbb{T}^d)}\,\mathrm{d}s\\
&\hspace{.5in}+\int_0^t\left\lVert \bu_n-\bu\right\rVert_{L^2(\mathbb{T}^d)}\left\lVert \bu\right\rVert_{L^2(\mathbb{T}^d)}\,\mathrm{d}s\\
&\lesssim\left\lVert \bu_n\right\rVert_{L^r([0,T],L^p(\mathbb{T}^d))}\left\lVert \bu_n-\bu\right\rVert_{L^r([0,T],L^p(\mathbb{T}^d))}\\
&\hspace{.5in}+\left\lVert \bu-\bu_n\right\rVert_{L^r([0,T],L^p(\mathbb{T}^d))}\left\lVert \bu\right\rVert_{L^r([0,T],L^p(\mathbb{T}^d))}\end{align*}
for $r,p>2.$ Recalling $\mathcal{Z}^{\sigma,\beta}_{r,p}\subseteq L^r([0,T],L^p(\mathbb{T}^d)),$ the expression above vanishes in the limit of $n\to\infty$.
The last term of (\ref{bieu}) is constant in $\bu$ and so trivially continuous. \\

\noindent 
(ii) Linearity is obvious. Recall that $\bph_n\to\bph$ in ${\mathcal S}({\mathbb R}^d)$ if the sequence of functions and all of their derivatives converge uniformly 
on ${\mathbb T}^d.$ Convergence of the first two terms of (\ref{bieu}) is immediate from the fact that $B_{p/2,\infty}^{\sigma-\beta}(\mathbb{T}^d)
\subset {\mathcal S}'(\mathbb{T}^d).$ For the third term, similarly as in (i), we note that 
\begin{align*}
\left|\int_0^t\left\langle\mathscr{P}\grad\cdot(\bu\otimes \bu)(\cdot,s),\bph-\bph_n\right\rangle\,\mathrm{d}s\right|&
=\left|\int_0^t\left\langle \bu\otimes \bu,\grad\mathscr{P}(\bph-\bph_n)\right\rangle\,\mathrm{d}s\right|\\
&\leq \int_0^t |\bu\otimes \bu|\,\mathrm{d}s\cdot \|\grad\mathscr{P}(\bph-\bph_n)\|_\infty\\
&\leq \|\bu\|^2_{\mathcal{Z}^{\sigma,\beta}_{r,p}}\|\grad\mathscr{P}(\bph-\bph_n)\|_\infty
\end{align*}
so that this term vanishes as $n\to\infty.$ Given the regularity $\bsf\in C^0([0,T]\times \mathbb{T}^d),$ the 
argument for the fourth term is very similar to that above. \\

\noindent 
(iii) Because $\bu\in C^0([0,T],B_{p/2,\infty}^{\sigma-\beta}(\mathbb{T}^d))$ and $\bph\in \mathcal{S}(\mathbb{T}^d)\subseteq B_{2/(2-p),1}^{\beta-\sigma}(\mathbb{T}^d),$
continuity of $\left\langle \bu(\cdot,t),\bph\right\rangle$ in $t$ is immediate. Regarding the time-integral terms, we note from the 
preceding estimates that the integrand functions $\langle(\bu\otimes \bu)(\cdot,s),\grad\mathscr{P}\bph\rangle$ and 
$\langle\bsf(\cdot,s),\mathscr{P}\bph\rangle$  are $L^1$ and the integrals are even absolutely continuous. \\  

\noindent 
(iv) The following simple estimates 
\[ |\left\langle \bu(\cdot,s),\bph\right\rangle|\lesssim\left\lVert \bu(\cdot,s)\right\rVert_{B_{p/2,\infty}^{\sigma-\beta}(\mathbb{T}^d)}\leq\left\lVert \bu\right\rVert_{C^0([0,T],B_{p/2,\infty}^{\sigma-\beta}(\mathbb{T}^d))},\quad s=0,t\hspace{0.15in}\]
\[\left|\int_0^t \left\langle\mathscr{P}\grad\cdot(\bu\otimes \bu),\bph\right\rangle\,\mathrm{d}s\right|=\left|\int_0^t\left\langle \bu\otimes \bu,\grad\mathscr{P}\bph\right\rangle\,\mathrm{d}s\right|\lesssim\int_0^t \left\lVert \bu\right\rVert^2_{L^2(\mathbb{T}^d)}\,\mathrm{d}s\lesssim\left\lVert \bu\right\rVert^2_{L^r([0,T],L^p(\mathbb{T}^d))}.\]
explain the terms linear and quadratic in $\|\bu\|_{\mathcal{Z}^{\sigma,\beta}_{r,p}}.$ The other expressions in (\ref{bieu}) yield the constant term. $\qed$

\hfill

\hfill 

Let us denote by $C^0_{b,B}(\mathcal{Z}^{\sigma,\beta}_{r,p},\mathbb{R})$ the set of bounded, continuous functions on $\mathcal{Z}^{\sigma,\beta}_{r,p}$
whose support in addition is bounded in the $\mathcal{Z}^{\sigma,\beta}_{r,p}$ norm.  For any $\eta\in C^0_{b,B}(\mathcal{Z}^{\sigma,\beta}_{r,p},\mathbb{R}),$ the mapping $\bu\mapsto E_t^\bsf(\bu,\bph)\eta(\bu)$ is $C^0_{b,B}(\mathcal{Z}^{\sigma,\beta}_{r,p},\mathbb{R})$ because of Lemma \ref{claim6}\,(i)\& (iv).  
Thus from part (i) of the Main Theorem, we have
\[\int E_t^\bsf(\bu,\bph)\eta(\bu)\,\mathrm{d}\mpr^\nu_\mu(\bu)\to\int E_t^\bsf(\bu,\bph)\eta(\bu)\,\mathrm{d}\mpr_\mu(\bu)\]
as $\nu\to0$. The strategy is to show that the sequence on the left in fact converges to 0, thus yielding 
$$ \int E_t^\bsf(\bu,\bph)\eta(\bu)\,\mathrm{d}\mpr_\mu(\bu) = 0. $$
This latter condition for all $\eta\in C^0_{b,B}(\mathcal{Z}^{\sigma,\beta}_{r,p},\mathbb{R})$ will then be shown to imply that 
$$  E_t^\bsf(\bu,\bph) =0 \quad \mpr_\mu-{\rm a.e.}\ \bu $$
These two statements are proved in the next two lemmas. 

\begin{lemma}\label{claim7} 
For any sequence $\nu\to 0$ and for any fixed $t\in [0,T],$ $\bph\in {\mathcal S}({\mathbb R}^d),$ $\eta\in C^0_{b,B}(\mathcal{Z}^{\sigma,\beta}_{r,p},\mathbb{R})$
$$ \lim_{\nu\to 0} \int E_t^\bsf(\bu,\bph)\eta(\bu)\,\mathrm{d}\mpr^\nu_\mu(\bu)=0, \quad $$
\end{lemma} 

\textit{Proof.} The proof is based on the definition of $P^\nu_\mu$
$$ \int E_t^\bsf(\bu,\bph)\eta(\bu)\,\mathrm{d}\mpr^\nu_\mu(\bu) = {\mathbb E}_\mu\left[E_t^\bsf(\bu^\nu,\bph)\eta(\bu^\nu)\right]$$
and the following expression for the Euler residual of solutions $\bu^\nu$ of \eqref{llns}:  
\begin{align}\label{bill}\begin{split}
E_t^\bsf(\bu^\nu,\bph)&=\int_0^t\left\langle\mathscr{P}\grad\cdot\left[(\bu^\nu\otimes \bu^\nu)- \pi_\Lambda(\bu^\nu\otimes \bu^\nu)\right],\bph\right\rangle \,\mathrm{d}s
+\int_0^t\nu\left\langle\triangle \pi_\Lambda \bu^\nu,\bph\right\rangle\,\mathrm{d}s\\
&\hspace{1in}+\nu^{\kappa}\int_0^t\left\langle\mathscr{P}\grad\cdot\mathrm{d} \pi_\Lambda \bxx(\cdot,s),\bph\right\rangle\mathrm{d}s
-\int_0^t\langle \bsf- \pi_\Lambda \bsf,\bph\rangle\,\mathrm{d}s.\end{split}\end{align}
In fact, the function $\eta\in C^0_{b,B}(\mathcal{Z}^{\sigma,\beta}_{r,p},\mathbb{R})$ plays no essential role in any estimates, aside from its boundedness. 

For the first term on the RHS of (\ref{bill}) we have
\begin{align*}\left|\int_0^t\left\langle\mathscr{P}\grad\cdot\left[(\bu^\nu\otimes \bu^\nu)- \pi_\Lambda(\bu^\nu\otimes \bu^\nu)\right],\bph\right\rangle\,\mathrm{d}s\right|&=
\left|\int_0^t\left\langle(\bu^\nu\otimes \bu^\nu)- \pi_\Lambda(\bu^\nu\otimes \bu^\nu),\grad\mathscr{P}\bph\right\rangle\,\mathrm{d}s\right|\\
&\lesssim\left\lVert(\bu^\nu\otimes^\nu)- \pi_\Lambda(\bu^\nu\otimes \bu^\nu)\right\rVert_{L^{r/2}([0,T],L^{p/2}(\mathbb{T}^d))}.\end{align*}
By the Lizorkin representation
\bea 
\left\lVert(\bu\otimes \bu)- \pi_\Lambda(\bu\otimes \bu)\right\rVert_{L^{r/2}([0,T],L^{p/2}(\mathbb{T}^d))}&\leq &
2^{-\sigma m(\nu)}\left\lVert\bu\otimes \bu\right\rVert_{L^{r/2}([0,T],B^{\sigma}_{p/2,\infty}(\mathbb{T}^d))} \cr
& \lesssim & 2^{-\sigma m(\nu)} \left\lVert \bu\right\rVert_{L^r([0,T],B_{p,\infty}^{\sigma}(\mathbb{T}^d))}^2
\eea  
This is sufficient to get
\bea
\left|{\mathbb E}_\mu\left[\eta(\bu^\nu)\int_0^t\left\langle\mathscr{P}\grad\cdot\left[(\bu^\nu\otimes \bu^\nu)- \pi_\Lambda(\bu^\nu\otimes \bu^\nu)\right],\bph\right\rangle\,\mathrm{d}s\,\right]\right| 
&\lesssim& 2^{-\sigma m(\nu)} \int \left\lVert \bu\right\rVert_{L^r([0,T],B_{p,\infty}^{\sigma}(\mathbb{T}^d))}^2 \,\mathrm{d}\mpr^\nu_\mu(\bu)\cr
&\to 0&. 
\eea

\hfill

We can similarly write
\[\left|\int_0^t\left\langle\triangle \pi_\Lambda \bu^\nu,\bph\right\rangle\,\mathrm{d}s\right|
\lesssim\left\lVert \bu^\nu\right\rVert_{C^0([0,T],B_{p/2,\infty}^{\sigma-\beta}(\mathbb{T}^d))}\|\triangle \pi_\Lambda\bph\|_{B_{(p-2)/p,1}^{\beta-\sigma}(\mathbb{T}^d))}.\]
Thus for the second term on the RHS of (\ref{bill}) we have
\[ \left|{\mathbb E}_\mu\left[\eta(\bu^\nu)\int\int_0^t\left\langle \nu\triangle \pi_\Lambda \bu^\nu,\bph\right\rangle\,\mathrm{d}s\right]\right|\lesssim
\nu \int \left\lVert \bu\right\rVert_{C^0([0,T],B_{p/2,\infty}^{\sigma-\beta}(\mathbb{T}^d))} \mpr^\nu_\mu(\bu)\to0,\]
The integral on the righthand side is finite by the tightness estimate in part (i) of the theorem and the whole term vanishes due to the coefficient $\nu$.

\hfill

For the third term of (\ref{bill}) we have
$$ \int_0^t\left\langle\mathscr{P}\grad\cdot \pi_\Lambda \mathrm{d}\bxx(\cdot,s),\bph\right\rangle
= -\int_0^t \langle \pi_\Lambda\mathrm{d}\bxx(\cdot,s),\grad\mathscr{P}\bph\rangle $$
Of course, from Cauchy-Schwartz
\[ \left|{\mathbb E}\left[\int_0^t\langle \pi_\Lambda\mathrm{d}\bxx(\cdot,s),\grad\mathscr{P}\bph\rangle\right]\right|
\leq\left({\mathbb E}\left|\int_0^t|\langle \pi_\Lambda\mathrm{d}\bxx(\cdot,s),\grad\mathscr{P}\bph\rangle\right|^2\right)^{1/2}.\]
Then using the Fourier representation of $\bxx,$ a simple calculation by It$\bar{{\rm o}}$ isometry yields: 
\begin{align*}{\mathbb E}\left[\left|\int_0^t\langle \pi_\Lambda\mathrm{d}\bxx(\cdot,s),\grad\mathscr{P}\bph\rangle\right|^2\right]&
\leq{\mathbb E}\left[\left|\int_0^t \sum_{|\bk|_\infty<2^{m(\nu)}}\sum_{i,j,\ell=1}^d\mathscr{P}_{j\ell}(\bk)k_i(\mathcal{F}\bph)_\ell(\bk)\,\mathrm{d}\bxx^\bk_{ij}(s)\right|^2\right]\\
&=\sum_{|\bk|_\infty<2^{m(\nu)}}\int_0^t\left|\sum_{i,j,\ell=1}^d[\mathscr{P}_{j\ell}(\bk)k_i+\mathscr{P}_{i\ell}(\bk)k_j](\mathcal{F}\bph)_\ell(\bk)\right|^2\,\mathrm{d}s\\
&\lesssim\sum_{|\bk|_\infty<2^{m(\nu)}} |\bk|^2 \left|(\mathcal{F}\bph)(\bk)\right|^2\ \lesssim\lVert\grad\bph\rVert_{L^2(\mathbb{T}^d)}^2.\end{align*}
Thus evidently
\[{\mathbb E}_\mu \left[\eta(\bu^\nu)\nu^{\kappa}\int_0^t\langle\mathscr{P}\grad\cdot \pi_\Lambda\mathrm{d}\bxx(\cdot,s),\bph\rangle\right]\to0.\]
We note here that the factor $\nu^{\kappa}$ could be replaced by any expression vanishing as $\nu\to 0.$

\hfill 

For the last term of (\ref{bill}) write
\[\left|\int_0^t\langle \bsf-\pi_\Lambda\bsf,\bph\rangle\,\mathrm{d}s\right|=\left|\int_0^t\langle \bsf,\bph-\pi_\Lambda\bph\rangle\,\mathrm{d}s\right|\lesssim\lVert\bsf\rVert_{C^0([0,T],B_{p/2,\infty}^{\sigma-\beta}(\mathbb{T}^d))}\lVert\bph-\pi_\Lambda\bph\rVert_{B_{2/(2-p),1}^{\sigma-\beta}(\mathbb{T}^d)}.\]
Since $\bph\in\mathcal{S}(\mathbb{T}^d)\subseteq B_{2/(2-p),1}^{\sigma-\beta}(\mathbb{T}^d)$, the last term vanishes as $\Lambda\to\infty$. $\qed$

\hfill

\begin{lemma}\label{claim8} 
Fix $t\in [0,T],$ $\bph\in {\mathcal S}({\mathbb R}^d).$ If 
$$\int E_t^\bsf(\bu,\bph)\eta(\bu)\,\mathrm{d}\mpr_\mu(\bu) = 0, \quad \forall \eta\in C^0_{b,B}(\mathcal{Z}^{\sigma,\beta}_{r,p},\mathbb{R}),$$
then 
$$  E_t^\bsf(\bu,\bph) =0 \quad \mpr_\mu-{\rm a.e.}\ \bu $$
\end{lemma} 

\textit{Proof.}
We define a sequence of functions $\chi_k\in C_{b,B}^0(\mathcal{Z}^{\sigma,\beta}_{r,p},[0,1])$ with $\chi_k(\bu)\uparrow 1$ 
as $k\uparrow\infty$ pointwise in $\bu$. First, we define the sets for $k\in {\mathbb N}^+$
$$ A_k = \{\bu\in \mathcal{Z}^{\sigma,\beta}_{r,p}: \|\bu\|_{\mathcal{Z}^{\sigma,\beta}_{r,p}}\leq k\}$$
$$ B_k = \{\bu\in \mathcal{Z}^{\sigma,\beta}_{r,p}: \|\bu\|_{\mathcal{Z}^{\sigma,\beta}_{r,p}}\geq k+1\}$$
which are clearly disjoint and closed in the strong topology of $\mathcal{Z}^{\sigma,\beta}_{r,p}.$ Since this topology for $\mathcal{Z}^{\sigma,\beta}_{r,p}$
is given by a metric, it is normal and the Urysohn lemma applies. Thus, for each $k\in {\mathbb N}^+$ there 
exists a function 
$\chi_k\in C^0(\mathcal{Z}^{\sigma,\beta}_{r,p},[0,1])$ such that $\left.\chi_k\right|_{A_k}=1$ and $\left.\chi_k\right|_{B_k}=0.$
Obviously, $\chi_k$ is supported in $A_{k+1},$ which is norm-bounded, and $\lim_{k\uparrow\infty}\chi_k\uparrow 1.$
Setting now $\eta_k(\bu):=E_t^\bsf(\bu,\bph)\chi_k(\bu)$, we observe that $\eta_k\in C_{b,B}^0(\mathcal{Z}^{\sigma,\beta}_{r,p}(\mathbb{T}^d))$
and thus we can apply Lemma \ref{claim7} to infer that
$$ \int\left|E_t^\bsf(\bu,\bph)\right|^2\chi_k(\bu)\,\mathrm{d}\mpr_\mu(\bu)= \int E_t^\bsf(\bu,\bph) \eta_k(\bu)\,\mathrm{d}\mpr_\mu(\bu)=0. $$
Taking the limit $k\to\infty,$ we obtain by monotone convergence that 
\[ \int\left|E_t^\bsf(\bu,\bph)\right|^2\,\mathrm{d}\mpr_\mu(\bu)=0. \]
It is immediate that $E_t^\bsf(\bu,\bph)=0$ for $\mpr_\mu-$a.e. $\bu.$ $\qed$ 

\hfill 

Using the previous lemmas, we now prove that realizations of $P_\mu$ are weak Euler solutions. For each 
$t\in [0,T],$ $\bph\in {\mathcal S}({\mathbb R}^d),$ we define the set  
\[W_{t,\bph}:=\left\{\bu\in \mathcal{Z}^{\sigma,\beta}_{r,p}:\, E_t^\bsf(\bu,\bph)=0\right\},\]
As $\bu\mapsto E_t^\bsf(\bu,\bph)$ is a continuous mapping $W_{t,\bph}$ is a closed set and thus measurable.  From our previous analysis, 
 $\mpr_\mu(W_{t,\bph})=1.$ In part (ii) of the theorem, we proved that $\mpr_\mu(\widetilde{\mathcal{E}}^{\sigma,\beta}_{r,p})=1,$ 
so that 
\[\mpr(W_{t,\bph}\cap \widetilde{\mathcal{E}}^{\sigma,\beta}_{r,p})=1.\]
We can thus replace $W_{t,\bph}$ with the measurable set $W_{t,\bph}\cap \widetilde{\mathcal{E}}^{\sigma,\beta}_{r,p},$
so that $W_{t,\bph}\subset  \widetilde{\mathcal{E}}^{\sigma,\beta}_{r,p}.$ 

Let $\mathscr{T}=[0,T]\cap {\mathbb Q}$ be the countable dense subset of rational times and denote by 
$\phi_\alpha(\bx)=e^{i\bk\bdot\bx} {\bf e}_i ,$ $\alpha\in \mathscr{A}=\{(\bk,i):\ \bk\in{\mathbb Z}^d,\, i\in \{1,2,3\}\},$
which by standard Fourier theory has dense linear span in $\mathcal{S}(\mathbb{T}^d)$. Defining
\[W_\alpha:=\bigcap_{t\in {\mathscr{T}}}W_{t,\bph_\alpha},\hspace{.15in}W:=\bigcap_{\alpha\in\mathscr{A}}W_\alpha.\]
we have $\mpr(W_\alpha)=1$ and $\mpr(W)=1,$ and furthermore 
$$ E_t^\bsf(\bu,\bph_\alpha)=0, \quad \forall t\in \mathscr{T}, \forall \alpha\in \mathscr{A}, \quad \forall \bu\in W. $$
Since $E_t^\bsf(\bu,\bph_\alpha)$ is continuous in $t$ for each $\alpha\in \mathscr{A}$ by Lemma \ref{claim6}(ii), we see that in fact 
$$ E_t^\bsf(\bu,\bph_\alpha)=0, \quad \forall t\in [0,T], \forall \alpha\in \mathscr{A}, \quad \forall \bu\in W. $$
Then, since $E_t^\bsf(\bu,\bph)$ is linear and continuous in $\bph$ by Lemma \ref{claim6}(iii) and since finite linear combinations 
of $\{\bph_\alpha:\, \alpha\in \mathscr{A}\}$
are dense in $\mathcal{S}(\mathbb{T}^d),$ then furthermore 
$$ E_t^\bsf(\bu,\bph)=0, \quad \forall t\in [0,T], \forall \bph\in \mathcal{S}(\mathbb{T}^d), \quad \forall \bu\in W. $$
We conclude that velocity fields $\mpr_\mu-$a.s. are mild Euler solutions in the sense of equation \eqref{mild} 

In fact, it is not hard to show that \eqref{mild} implies the standard notion of weak Euler solution. We sketch here 
the argument. Any $\psi\in\mathcal{S}(0,T)$ must be supported on some closed interval $[h_0,T-h_0]$ for $0<h_0<T/2.$
Smearing the equation \eqref{mild} with the derivative function $\psi'$ 
then gives 
\be  \int_0^T E_t^\bsf(\bu,\bph) \psi'(t)\, \mathrm{d}t=0. \lb{weak1} \ee
Note that the term $\left\langle \bu(\cdot,0),\bph\right\rangle$ in equation \eqref{bieu} for the Euler residual 
is constant in time and thus gives no contribution to the above integral. As noted in the proof of Lemma \ref{claim6},(iii),
the time-integral part of $E_t^\bsf(\bu,\bph)$ in equation \eqref{bieu} given by 
$$ K_t^\bsf(\bu,\bph):=\int_0^t\left\langle\mathscr{P}[\grad\cdot(\bu\otimes \bu)-\bsf](\cdot,s),\bph\right\rangle\,\mathrm{d}s $$
is in fact absolutely continuous in time $t.$ Thus, integration by parts is justified and one obtains
\bea \int_0^T K_t^\bsf(\bu,\bph) \psi'(t)\, \mathrm{d}t &=& - \int_0^T \frac{\mathrm{d}}{\mathrm{d}t}K_t^\bsf(\bu,\bph) \psi(t)\, \mathrm{d}t \cr
&=& -\int_0^T \left\langle\mathscr{P}[\grad\cdot(\bu\otimes \bu)-\bsf](\cdot,t),\bph\right\rangle \psi(t) \, \mathrm{d}t. \eea 
We thus obtain from the equation \eqref{weak1} the condition 
\be  \int_0^T \left\langle \bu(\cdot,t),\bph\right\rangle\psi'(t)\, \mathrm{d}t - \int_0^T \left\langle\mathscr{P}[\grad\cdot(\bu\otimes \bu)-\bsf](\cdot,t),\bph\right\rangle \psi(t) \, \mathrm{d}t=0. \lb{weak} \ee
which is essentially the standard weak formulation of the Euler equations. In view of part (iv) of the theorem concerning 
initial data, which we prove in the next subsection, it is useful to remark that this same argument may be applied even if 
$\psi(0)\neq 0.$ In that case, one obtains instead 
\be  \int_0^T \left\langle \bu(\cdot,t),\bph\right\rangle\psi'(t)\, \mathrm{d}t + \left\langle \bu(\cdot,0),\bph\right\rangle\psi(0)- \int_0^T \left\langle\mathscr{P}[\grad\cdot(\bu\otimes \bu)-\bsf](\cdot,t),\bph\right\rangle \psi(t) \, \mathrm{d}t=0. \lb{weak} \ee
This is a standard approach to incorporate initial data in the weak formulation of a PDE. 

Finally, we note that a weak formulation of the incompressibility condition follows directly from \eqref{mild}. 
If we take $\bph=\grad\chi$ for a scalar function $\chi\in {\mathcal S}({\mathbb T}^d)$, then substituting into \eqref{mild} 
it follows that all of the time-integral terms vanish in the equation \eqref{bieu} defining $E_t^\bsf(\bu,\bph)$, because of the 
Leray projection $\mathscr{P}.$ The resulting equation is 
$$ \left\langle \bu(\cdot,t),\grad\chi\right\rangle=\left\langle \bu(\cdot,0),\grad\chi\right\rangle, \quad
\forall t\in [0,T], \forall \chi\in \mathcal{S}(\mathbb{T}^d).
$$
Recall that initial data for the Landau-Lifschitz equation \eqref{llns} are chosen from the measure $\mu$ such that 
$$ \left\langle \bu_0,\grad\chi\right\rangle=0, \quad  \forall \chi\in \mathcal{S}(\mathbb{T}^d) 
\qquad \mu-{a.e.} \ \bu_0 $$  
Thus, we infer the weak incompressibility condition
$$ \left\langle \bu(\cdot,t),\grad\chi\right\rangle=0, \quad \forall t\in [0,T], \forall \chi\in \mathcal{S}(\mathbb{T}^d)
\qquad P_\mu-{a.e.} \ \bu$$
if it can be shown that the distribution of $\bu(\cdot,0)$ induced by the measure $P_\mu$ on $\bu$ coincides 
with measure $\mu.$ This is precisely the result established in the next section. 

\hfill

\subsection{Proof of (iv)}: Our proof is modeled on Theorem IV.5.3 of \cite{vishik1988mathematical}. We have already 
observed that the trace $\gamma_0\bu=\bu(\cdot, 0)$ is well-defined as a map 
$\gamma_0:C^0([0,T],B_{p/2,\infty}^{\sigma-\beta}(\mathbb{T}^d))\to B_{p/2,\infty}^{\sigma-\beta}(\mathbb{T}^d)$
and is strongly continuous on $\mathcal{Z}^{\sigma,\beta}_{r,p}.$ Thus, the pushforward measure $\mpr^{(0)}_\mu:
=\gamma_{0\ast}P_\mu$ is well-defined on $\mathcal{B}(B_{p/2,\infty}^{\sigma-\beta})$. Our strategy to show 
the equivalence of $\mpr^{(0)}_\mu$ and $\mu$ is based on the study of their characteristic functions: 
$$ S_\mu^{(0)}[\bv]=\int\exp\left(i\langle \bu_0,\mathbf{v}\rangle\right)\,\mathrm{d}\mpr _\nu^{(0)}(\bu_0), \quad
 S_\mu[\bv]=\int\exp\left(i\langle \bu_0,\mathbf{v}\rangle\right)\,\mathrm{d}\mu(\bu_0) $$
defined here for $\mathbf{v}\in\mathcal{S}(\mathbb{T}^d).$ We note first that one can define similarly 
$\mpr^{\nu,(0)}_\mu:=\gamma_{0\ast}P_\mu^\nu$ for $\nu>0$ and its 
characteristic function satisfies 
\[  S_\mu^{\nu,(0)}[\bv]:=\int\exp\left(i\langle \bu_0,\mathbf{v}\rangle\right)\,\mathrm{d}\mpr_\mu^{\nu,(0)}(\bu_0)
=\int\exp\left(i\langle\gamma_0\bu,\mathbf{v}\rangle\right)\,\mathrm{d}\mpr_\mu^{\nu}(\bu) \] 
Because $\exp\left(i\langle\gamma_0\bu,\mathbf{v}\rangle\right)$ is a bounded, continuous function, it follows
from weak convergence that 
\[\lim_{\nu\to 0} S_\mu^{\nu,(0)}[\bv]=\int\exp\left(i\langle\gamma_0\bu,\mathbf{v}\rangle\right)\,\mathrm{d}\mpr_\mu (\bu)=\int\exp\left(i\langle \bu_0,\mathbf{v}\rangle\right)\,\mathrm{d}\mpr^{(0)}_\mu(\bu_0)=S_\mu^{(0)}[\bv]. \]
On the other hand, we have also by the definition $\bu^\nu_0:=\pi_\Lambda\bu_0$ that 
\[ S_\mu^{\nu,(0)}[\bv]=\int\exp\left(i\langle \pi_\Lambda\bu_0,\mathbf{v}\rangle\right)\,\mathrm{d}\mu(\bu_0). \] 
Since $ \pi_\Lambda \bu_0\to \bu_0$ in $L^p(\mathbb{T}^d)$ and thus almost sure in $\mu$, 
dominated convergence gives also that 
\[ \lim_{\nu\to 0} S_\mu^{\nu,(0)}[\bv]=\int\exp\left(i\langle \bu_0,\bv\rangle\right)\,\mathrm{d}\mu(\bu_0)=S_\mu[\bv].\]
Comparing the two limit expressions, we conclude that
\be S_\mu^{(0)}[\bv]=S_\mu[\bv]. \lb{chareq} \ee 

We would like to infer from this equality that $\mpr^{(0)}_\mu=\mu$ by applying the Bochner--Minlos theorem \cite{glimm2012quantum}.
However, this theorem applies to measures on $\mathcal{B}(\mathcal{S}^\prime(\mathbb{T}^d)),$ whereas $\mpr^{(0)}_\mu$ is defined 
on $\mathcal{B}(B_{p/2,\infty}^{\sigma-\beta})$ and $\mu$ on $\mathcal{B}(L^p(\mathbb{T}^d)).$ To resolve this difficulty we must 
extend both of these measures to $\mathcal{B}(\mathcal{S}^\prime(\mathbb{T}^d)),$ which is the content of the following:  

\begin{lemma}\label{claim9} (i) The set $B_{p/2,\infty}^{\sigma-\beta}(\mathbb{T}^d)$ belongs to $\mathcal{B}(\mathcal{S}^\prime(\mathbb{T}^d))$
and the measure $\mpr^{(0)}_\mu$ has a unique extension $\overline{\mpr}^{(0)}_\mu$ to $\mathcal{B}(\mathcal{S}^\prime(\mathbb{T}^d))$ 
that is supported on $B_{p/2,\infty}^{\sigma-\beta}(\mathbb{T}^d),$ defined by 
\be \overline{\mpr}^{(0)}_\mu(E)=\mpr^{(0)}_\mu(E\cap B_{p/2,\infty}^{\sigma-\beta}(\mathbb{T}^d)), \quad \forall E\in\mathcal{B}(\mathcal{S}^\prime(\mathbb{T}^d)) \lb{Pex} \ee \\ 
(ii) The set $L^p(\mathbb{T}^d)$ belongs to $\mathcal{B}(\mathcal{S}^\prime(\mathbb{T}^d))$ and the measure $\mu$ has a unique 
extension $\overline{\mu}$ to $\mathcal{B}(\mathcal{S}^\prime(\mathbb{T}^d))$ that is supported on $L^p(\mathbb{T}^d),$ defined by 
\be \overline{\mu}(E)=\mu(E\cap L^p(\mathbb{T}^d)), \quad \forall E\in\mathcal{B}(\mathcal{S}^\prime(\mathbb{T}^d)). \lb{muex} \ee 
\end{lemma} 

\hfill 

\textit{Proof.} The argument is an elaboration of Remark II.2.1 in \cite{vishik1988mathematical}. (i) We prove Borel measurability of  $B_{p/2,\infty}^{\sigma-\beta}(\mathbb{T}^d)$ in $\mathcal{S}^\prime(\mathbb{T}^d).$
Note first that the extended real-valued function defined on  $\mathcal{S}^\prime(\mathbb{T}^d)$ by 
\[\bu\mapsto\lVert \bu\rVert_{B_{p/2,\infty}^{\sigma-\beta}(\mathbb{T}^d)}\]
is lower semicontinuous in the Fr\'echet topology of $\mathcal{S}^\prime(\mathbb{T}^d)$. To see this, 
take $\bu_n\to \bu$ in $\mathcal{S}^\prime(\mathbb{T}^d)$ and note 
\[\forall\bph\in\mathcal{S}(\mathbb{T}^d):|\langle \bu_n,\bph\rangle|\leq\lVert\bph\rVert_{B_{2/(2-p),1}^{\beta-\sigma}(\mathbb{T}^d)}\lVert \bu_n\rVert_{B_{p/2,\infty}^{\sigma-\beta}(\mathbb{T}^d)}.\]
Taking the limit infimum as $n\to\infty$ of both sides of this inequality gives
\[\forall\bph\in\mathcal{S}(\mathbb{T}^d):|\langle \bu,\bph\rangle|\leq\lVert\bph\rVert_{B_{2/(2-p),1}^{\beta-\sigma}(\mathbb{T}^d)}\liminf_{n\to\infty}\lVert \bu_n\rVert_{B_{p/2,\infty}^{\sigma-\beta}(\mathbb{T}^d)}.\]
Then, using the dense embedding $\mathcal{S}(\mathbb{T}^d)\hookrightarrow B_{2/(2-p),1}^{\beta-\sigma}(\mathbb{T}^d),$
\[\lVert \bu\rVert_{B_{p/2,\infty}^{\sigma-\beta}(\mathbb{T}^d)}\equiv\sup_{\bph\in\mathcal{S}(\mathbb{T}^d),\, \lVert\bph\rVert_{B_{2/(2-p),1}^{\beta-\sigma}(\mathbb{T}^d)}=1\}}|\langle \bu,\bph\rangle|\leq\liminf_{n\to\infty}\lVert \bu_n\rVert_{B_{p/2,\infty}^{\sigma-\beta}(\mathbb{T}^d)},\]
which proves lower semicontinuity. As a consequence, the sets
\[M^{\sigma,\beta}_{p}(n):=\left\{\bu\in\mathcal{S}^\prime(\mathbb{T}^d):\lVert \bu\rVert_{B_{p/2,\infty}^{\sigma-\beta}(\mathbb{T}^d)}\leq n\right\}\]
are closed in the Fr\'echet topology of $\mathcal{S}^\prime(\mathbb{T}^d)$ and then 
\[B_{p/2,\infty}^{\sigma-\beta}(\mathbb{T}^d)\equiv\bigcup_{n\in\mathbb{N}}M^{\sigma,\beta}_{p}(n)\]
is measurable in $\mathcal{B}(\mathcal{S}^\prime(\mathbb{T}^d))$. In fact, we have proved that 
$B_{p/2,\infty}^{\sigma-\beta}(\mathbb{T}^d)$ is an $F^\sigma$-set in $\mathcal{S}^\prime(\mathbb{T}^d).$

A measure $ \overline{\mpr}^{(0)}_\mu$ defined on $\mathcal{B}(\mathcal{S}^\prime(\mathbb{T}^d))$ that is supported on 
the measurable set $B_{p/2,\infty}^{\sigma-\beta}(\mathbb{T}^d)$ must satisfy  
$$ \overline{\mpr}^{(0)}_\mu(E)=\overline{\mpr}^{(0)}_\mu(E\cap B_{p/2,\infty}^{\sigma-\beta}(\mathbb{T}^d)), \quad \forall E\in\mathcal{B}(\mathcal{S}^\prime(\mathbb{T}^d)). $$
If furthermore this measure restricted to $B_{p/2,\infty}^{\sigma-\beta}(\mathbb{T}^d)$ coincides with $\mpr^{(0)}_\mu,$ then it must hold that 
$$ \overline{\mpr}^{(0)}_\mu(E)=\mpr^{(0)}_\mu(E\cap B_{p/2,\infty}^{\sigma-\beta}(\mathbb{T}^d)), \quad \forall E\in\mathcal{B}(\mathcal{S}^\prime(\mathbb{T}^d)), $$
which is formula \eqref{Pex}. This establishes uniqueness.

However, to show existence, one must show that 
\begin{equation}\label{borelbesov}\left\{E\cap B_{p/2,\infty}^{\sigma-\beta}(\mathbb{T}^d):E\in\mathcal{B}(\mathcal{S}^\prime(\mathbb{T}^d))\right\}\subseteq\mathcal{B}(B_{p/2,\infty}^{\sigma-\beta}(\mathbb{T}^d)), \end{equation}
so that the formula \eqref{Pex} is meaningful. To prove this, recall that the topology on $\mathcal{S}^\prime(\mathbb{T}^d)$ 
is generated by the basis of open sets of the form
\be   \{ \bar{\Lambda}_\bph^{-1}(U):\, \bph\in {\mathcal S}({\mathbb T}^d),\ {\rm open}\ U\subset {\mathbb R}\}  \lb{basis} \ee
where $\bar{\Lambda}_\bph: \mathcal{S}^\prime(\mathbb{T}^d)\to {\mathbb R}$ for $\bph\in {\mathcal S}({\mathbb T}^d)$
is defined by $\bar{\Lambda}_\bph(\bv):=
\langle \bv,\bph\rangle$ for $\bv\in \mathcal{S}^\prime(\mathbb{T}^d).$ Clearly, one can also define a corresponding functional 
$\Lambda_\bph: B_{p/2,\infty}^{\sigma-\beta}(\mathbb{T}^d)\to {\mathbb R}$ by the same formula $\Lambda_\bph(\bv):=
\langle \bv,\bph\rangle$ for $\bv\in B_{p/2,\infty}^{\sigma-\beta}(\mathbb{T}^d)$. It is then easy to check that
$$ \bar{\Lambda}_\bph^{-1}(U) \cap B_{p/2,\infty}^{\sigma-\beta}(\mathbb{T}^d) = \Lambda_\bph^{-1}(U). $$ 
However, because ${\mathcal S}({\mathbb T}^d)\subset B_{(p-2)/p,1}^{\beta-\sigma}(\mathbb{T}^d),$ the function
$\Lambda_\bph$ is strongly continuous and thus the set $\Lambda_\bph^{-1}(U)$ is open in $B_{p/2,\infty}^{\sigma-\beta}(\mathbb{T}^d).$ 
Since the Borel $\sigma$-algebra $\mathcal{B}(\mathcal{S}^\prime(\mathbb{T}^d))$ is generated by the basis \eqref{basis}, the 
requirement \eqref{borelbesov} for existence follows by standard Boolean set algebra.

(i) The proofs of the corresponding statements for $L^p({\mathbb R}^d)$ and the measure $\mu$ are almost identical and need not be repeated here. $\qed$

\hfill

In order to apply this lemma, note that the function $\bu_0\mapsto \langle \bu_0,\bv\rangle $ for $\bv\in {\mathcal S}({\mathbb R}^d)$ extends 
to a measurable function on $\mathcal{S}^\prime(\mathbb{T}^d).$ Thus, we can rewrite the equality \eqref{chareq} instead as the equality 
\be \overline{S}_\mu^{(0)}[\bv]=\overline{S}_\mu[\bv] \lb{chareq2} \ee
for the characteristic functions of the measures $\overline{\mpr}^{(0)}_\mu$ and $\overline{\mu}$ defined on $\mathcal{B}(\mathcal{S}^\prime(\mathbb{T}^d)).$
Finally, the Bochner-Minlos theorem implies that 
$$ \overline{\mpr}^{(0)}_\mu=\overline{\mu}. $$ 
It follows from this equality that $\overline{\mpr}^{(0)}_\mu$ is in fact supported on $L^p(\mathbb{T}^d)$. Now, we know that $\overline{\mpr}^{(0)}_\mu$
restricted to $\mathcal{B}(B_{p/2,\infty}^{\sigma-\beta}(\mathbb{T}^d))$ coincides with $\mpr^{(0)}_\mu$. Evidently $L^p(\mathbb{T}^d)$ is reflexive and separable, and per Lemma~\ref{claim2} embeds continuously into $B_{p/2,\infty}^{\sigma-\beta}(\mathbb{T}^d)$. Thus, by Theorem II.2.1 and Lemma II.2.1 of \cite{vishik1988mathematical}, 
\be \mathcal{B}(B_{p/2,\infty}^{\sigma-\beta}(\mathbb{T}^d))\cap L^p(\mathbb{T}^d)= \mathcal{B}(L^p(\mathbb{T}^d)). 
\lb{boreleq} \ee
Note that the statement of Lemma II.2.1 in \cite{vishik1988mathematical} required that both spaces $\Omega_1,$ $\Omega_2$
be separable, whereas in our case $\Omega_2=B_{p/2,\infty}^{\sigma-\beta}(\mathbb{T}^d)$ is non-separable. 
Separability and reflexivity of $\Omega_1$ were both needed in the proof of  \cite{vishik1988mathematical}, Lemma II.2.1
to extract a weakly convergent subsequence, but neither property was required of $\Omega_2.$
For our case $\Omega_1=L^p(\mathbb{T}^d)$ is both separable and reflexive. The equality \eqref{boreleq} thus holds 
and we can further restrict $\mpr^{(0)}_\mu$ to $\mathcal{B}(L^p(\mathbb{T}^d)),$ which must then coincide with $\overline{\mu}$ restricted to $\mathcal{B}(L^p(\mathbb{T}^d)),$ which is $\mu.$

\section*{Acknowledgements} 
\noindent 
We are grateful to Dmytro Bandak, Theodore Drivas, Nigel Goldenfeld, and Alexei Mailybaev for discussions 
of this problem. This work was funded by the Simons Foundation, via Targeted Grant in MPS-663054 
and the Collaboration Grant No. MPS- 1151713.

%
%

\bibliographystyle{unsrt}
\bibliography{bibliography.bib}

\begin{thebibliography}{10}

\bibitem{falkovich2001particles}
Gregory Falkovich, K~Gaw\c{e}dzki, and Massimo Vergassola.
\newblock Particles and fields in fluid turbulence.
\newblock {\em Reviews of modern Physics}, 73(4):913, 2001.

\bibitem{richardson1926atmospheric}
Lewis~Fry Richardson.
\newblock Atmospheric diffusion shown on a distance-neighbour graph.
\newblock {\em Proceedings of the Royal Society of London. Series A},
  110(756):709--737, 1926.

\bibitem{bernard1998slow}
Denis Bernard, Krzysztof Gawedzki, and Antti Kupiainen.
\newblock Slow modes in passive advection.
\newblock {\em Journal of Statistical Physics}, 90(3):519--569, 1998.

\bibitem{onsager1949statistical}
Lars Onsager.
\newblock Statistical hydrodynamics.
\newblock {\em Il Nuovo Cimento (1943-1954)}, 6(2):279--287, 1949.

\bibitem{kraichnan1968small}
Robert~H Kraichnan.
\newblock Small-scale structure of a scalar field convected by turbulence.
\newblock {\em The Physics of Fluids}, 11(5):945--953, 1968.

\bibitem{drivas2017lagrangian}
Theodore~D Drivas and Gregory~L Eyink.
\newblock A lagrangian fluctuation--dissipation relation for scalar turbulence.
  part i. flows with no bounding walls.
\newblock {\em Journal of Fluid Mechanics}, 829:153--189, 2017.

\bibitem{mailybaev2016spontaneously}
Alexei~A Mailybaev.
\newblock Spontaneously stochastic solutions in one-dimensional inviscid
  systems.
\newblock {\em Nonlinearity}, 29(8):2238, 2016.

\bibitem{mailybaev2016spontaneous}
Alexei~A Mailybaev.
\newblock Spontaneous stochasticity of velocity in turbulence models.
\newblock {\em Multiscale Modeling \& Simulation}, 14(1):96--112, 2016.

\bibitem{lorenz1969predictability}
Edward~N Lorenz.
\newblock The predictability of a flow which possesses many scales of motion.
\newblock {\em Tellus}, 21(3):289--307, 1969.

\bibitem{de2010admissibility}
Camillo De~Lellis and L{\'a}szl{\'o} Sz{\'e}kelyhidi.
\newblock On admissibility criteria for weak solutions of the euler equations.
\newblock {\em Archive for rational mechanics and analysis}, 195(1):225--260,
  2010.

\bibitem{de2017high}
Camillo De~Lellis and L{\'a}szl{\'o} Sz{\'e}kelyhidi~Jr.
\newblock High dimensionality and h-principle in pde.
\newblock {\em Bulletin of the American Mathematical Society}, 54(2):247--282,
  2017.

\bibitem{daneri2021non}
Sara Daneri, Eris Runa, and L{\'a}szl{\'o} Sz{\'e}kelyhidi.
\newblock Non-uniqueness for the euler equations up to onsager’s critical
  exponent.
\newblock {\em Annals of PDE}, 7(1):1--44, 2021.

\bibitem{szekelyhidi2011weak}
L{\'a}szl{\'o} Sz{\'e}kelyhidi~Jr.
\newblock Weak solutions to the incompressible euler equations with vortex
  sheet initial data.
\newblock {\em Comptes Rendus Mathematique}, 349(19-20):1063--1066, 2011.

\bibitem{thalabard2020butterfly}
Simon Thalabard, J{\'e}r{\'e}mie Bec, and Alexei~A Mailybaev.
\newblock From the butterfly effect to spontaneous stochasticity in singular
  shear flows.
\newblock {\em Communications Physics}, 3(1):1--8, 2020.

\bibitem{palmer2014real}
TN~Palmer, A~D{\"o}ring, and G~Seregin.
\newblock The real butterfly effect.
\newblock {\em Nonlinearity}, 27(9):R123, 2014.

\bibitem{betchov1957fine}
R~Betchov.
\newblock On the fine structure of turbulent flows.
\newblock {\em Journal of Fluid Mechanics}, 3(2):205--216, 1957.

\bibitem{betchov1961thermal}
R.~Betchov.
\newblock Thermal agitation and turbulence.
\newblock In L.~Talbot, editor, {\em Rarefied Gas Dynamics}, page 307–321.
  Academic Press, New York, 1961.
\newblock Proceedings of the Second International Symposium on Rarefied Gas
  Dynamics, held at the University of California, Berkeley, CA, 1960.

\bibitem{ruelle1979microscopic}
David Ruelle.
\newblock Microscopic fluctuations and turbulence.
\newblock {\em Physics Letters A}, 72(2):81--82, 1979.

\bibitem{landau1959fluid}
L.D. Landau and E.~M. Lifschitz.
\newblock {\em Fluid Mechanics}.
\newblock Pergamon Press, 1959.

\bibitem{forster1977large}
Dieter Forster, David~R Nelson, and Michael~J Stephen.
\newblock Large-distance and long-time properties of a randomly stirred fluid.
\newblock {\em Physical Review A}, 16(2):732, 1977.

\bibitem{donev2014low}
Aleksandar Donev, Andy Nonaka, Yifei Sun, Thomas Fai, Alejandro Garcia, and
  John Bell.
\newblock Low mach number fluctuating hydrodynamics of diffusively mixing
  fluids.
\newblock {\em Communications in Applied Mathematics and Computational
  Science}, 9(1):47--105, 2014.

\bibitem{schwenk2012renormalization}
A.~Schwenk and J.~Polonyi.
\newblock {\em Renormalization Group and Effective Field Theory Approaches to
  Many-Body Systems}, volume 852 of {\em Lecture Notes in Physics}.
\newblock Springer, Berlin Heidelberg, 2012.

\bibitem{liu2018lectures}
Hong Liu and Paolo Glorioso.
\newblock Lectures on non-equilibrium effective field theories and fluctuating
  hydrodynamics.
\newblock In {\em Theoretical Advanced Study Institute Summer School 2017"
  Physics at the Fundamental Frontier"}, volume 305, page 008. Sissa Medialab,
  2018.

\bibitem{gallis2021turbulence}
MA~Gallis, JR~Torczynski, MC~Krygier, NP~Bitter, and SJ~Plimpton.
\newblock Turbulence at the edge of continuum.
\newblock {\em Physical Review Fluids}, 6(1):013401, 2021.

\bibitem{bandak2023spontaneous}
Dmytro Bandak, Alexei Mailybaev, Gregory~L. Eyink, and Nigel Goldenfeld.
\newblock Spontaneous stochasticity amplifies even thermal noise to the largest
  scales of turbulence in a few eddy turnover times.
\newblock {\em Phys. Rev. Lett.}, 132:104002, 2024.

\bibitem{eyink2020renormalization}
Gregory~L Eyink and Dmytro Bandak.
\newblock Renormalization group approach to spontaneous stochasticity.
\newblock {\em Physical Review Research}, 2(4):043161, 2020.

\bibitem{mailybaev2022spontaneous}
Alexei~A Mailybaev and Artem Raibekas.
\newblock Spontaneous stochasticity and renormalization group in discrete
  multi-scale dynamics.
\newblock {\em Communications in Mathematical Physics}, 401(3):2643--2671,
  2023.

\bibitem{vishik1988mathematical}
M.I. Vishik and A.V. Fursikov.
\newblock {\em Mathematical Problems of Statistical Hydromechanics}, volume~54
  of {\em Mathematics and its Applications}.
\newblock Kluwer Academic Publishers, Netherlands, 1988.

\bibitem{lions1996mathematical}
P.L. Lions.
\newblock {\em Mathematical Topics in Fluid Mechanics: Volume 1: Incompressible
  Models}, volume~3 of {\em Oxford Lecture Series in Mathematics and its
  Applications}.
\newblock Clarendon Press, Oxford, 1996.

\bibitem{diperna1987oscillations}
Ronald~J DiPerna and Andrew~J Majda.
\newblock Oscillations and concentrations in weak solutions of the
  incompressible fluid equations.
\newblock {\em Communications in mathematical physics}, 108(4):667--689, 1987.

\bibitem{fjordholm2016computation}
Ulrik~S Fjordholm, Siddhartha Mishra, and Eitan Tadmor.
\newblock On the computation of measure-valued solutions.
\newblock {\em Acta numerica}, 25:567--679, 2016.

\bibitem{thalabard2020turbulence}
Simon Thalabard and J{\'e}r{\'e}mie Bec.
\newblock Turbulence of generalised flows in two dimensions.
\newblock {\em Journal of Fluid Mechanics}, 883:A49, 2020.

\bibitem{chen2012kolmogorov}
Gui-Qiang Chen and James Glimm.
\newblock Kolmogorov's theory of turbulence and inviscid limit of the
  {N}avier-{S}tokes equations in {${\mathbb R}^3$}.
\newblock {\em Commun. Math. Phys.}, 310(1):267--283, 2012.

\bibitem{constantin2018remarks}
Peter Constantin and Vlad Vicol.
\newblock Remarks on high reynolds numbers hydrodynamics and the inviscid
  limit.
\newblock {\em Journal of Nonlinear Science}, 28:711--724, 2018.

\bibitem{drivas2019onsager}
Theodore~D Drivas and Gregory~L Eyink.
\newblock An onsager singularity theorem for leray solutions of incompressible
  navier--stokes.
\newblock {\em Nonlinearity}, 32(11):4465, 2019.

\bibitem{drivas2019remarks}
Theodore~D Drivas and Huy~Q Nguyen.
\newblock Remarks on the emergence of weak euler solutions in the vanishing
  viscosity limit.
\newblock {\em Journal of Nonlinear Science}, 29(2):709--721, 2019.

\bibitem{isett2022nonuniqueness}
Philip Isett.
\newblock Nonuniqueness and existence of continuous, globally dissipative euler
  flows.
\newblock {\em Archive for Rational Mechanics and Analysis}, 244(3):1223--1309,
  2022.

\bibitem{fjordholm2024vanishing}
Ulrik~Skre Fjordholm, Siddhartha Mishra, and Franziska Weber.
\newblock On the vanishing viscosity limit of statistical solutions of the
  incompressible navier--stokes equations.
\newblock {\em SIAM Journal on Mathematical Analysis}, 56(4):5099--5143, 2024.

\bibitem{reichelsdorfer2016foundations}
Martin Reichelsdorfer.
\newblock {\em Foundations of Small Scale Hydrodynamics with External Friction
  and Slip}.
\newblock PhD thesis, Friedrich-Alexander Universit\"{a}t
  Erlangen-N\"{u}rnberg,
  \url{https://opus4.kobv.de/opus4-fau/frontdoor/index/index/docId/7034}, 2016.

\bibitem{brehier2014short}
Charles-Edouard Br{\'e}hier.
\newblock A short introduction to stochastic pdes.
\newblock \url{https://hal.science/hal-00973887v2}, 2014.

\bibitem{flandoli2008introduction}
F.~Flandoli.
\newblock An introduction to 3d stochastic fluid dynamics.
\newblock In G.~Da~Prato and M.~R{\"o}ckner, editors, {\em SPDE in
  Hydrodynamics: Recent Progress and Prospects: Lectures given at the C.I.M.E.
  Summer School held in Cetraro, Italy, August 29 - September 3, 2005}, volume
  1942 of {\em Lecture Notes in Mathematics}. Springer, Berlin Heidelberg,
  2008.

\bibitem{drivas2022self}
Theodore~D Drivas.
\newblock Self-regularization in turbulence from the kolmogorov 4/5-law and
  alignment.
\newblock {\em Philosophical Transactions of the Royal Society A},
  380(2226):20210033, 2022.

\bibitem{constantin2007regularity}
Peter Constantin, Boris Levant, and Edriss~S Titi.
\newblock Regularity of inviscid shell models of turbulence.
\newblock {\em Physical Review E—Statistical, Nonlinear, and Soft Matter
  Physics}, 75(1):016304, 2007.

\bibitem{mailybaev2023spontaneous}
Alexei~A Mailybaev and Artem Raibekas.
\newblock Spontaneous stochasticity and renormalization group in discrete
  multi-scale dynamics.
\newblock {\em Commun. Math. Phys.}, 401:2643--2671, 2023.

\bibitem{mailybaev2023spontaneously}
Alexei~A Mailybaev and Artem Raibekas.
\newblock Spontaneously stochastic arnold’s cat.
\newblock {\em Arnold Mathematical Journal}, 9(3):339--357, 2023.

\bibitem{bedrossian2022regularity}
Jacob Bedrossian, Alex Blumenthal, and Sam Punshon-Smith.
\newblock A regularity method for lower bounds on the lyapunov exponent for
  stochastic differential equations.
\newblock {\em Inventiones mathematicae}, 227(2):429--516, 2022.

\bibitem{bedrossian2024chaos}
Jacob Bedrossian and Sam Punshon-Smith.
\newblock Chaos in stochastic 2d galerkin-navier--stokes.
\newblock {\em Communications in Mathematical Physics}, 405(4):107, 2024.

\bibitem{mailybaev2024rg}
Alexei~A Mailybaev.
\newblock Rg approach to the inviscid limit for shell models of turbulence.
\newblock {\em arXiv preprint arXiv:2408.04659}, 2024.

\bibitem{duchon2000inertial}
Jean Duchon and Raoul Robert.
\newblock Inertial energy dissipation for weak solutions of incompressible
  euler and navier-stokes equations.
\newblock {\em Nonlinearity}, 13(1):249, 2000.

\bibitem{eyink2007turbulenceI}
Gregory~L Eyink.
\newblock 2021, {T}urbulence {T}heory {I}. {C}ourse {N}otes, {T}he {J}ohns
  {H}opkins {U}niversity.
\newblock \url{http://www.ams.jhu.edu/~eyink/Turbulence/notes}, 2007.

\bibitem{gess2023landau}
Benjamin Gess, Daniel Heydecker, and Zhengyan Wu.
\newblock Landau-lifshitz-navier-stokes equations: Large deviations and
  relationship to the energy equality.
\newblock {\em arXiv preprint arXiv:2311.02223}, 2023.

\bibitem{bardos1991fluid}
Claude Bardos, Fran{\c{c}}ois Golse, and David Levermore.
\newblock Fluid dynamic limits of kinetic equations. i. formal derivations.
\newblock {\em Journal of statistical physics}, 63:323--344, 1991.

\bibitem{bardos1993fluid}
Claude Bardos, Fran{\c{c}}ois Golse, and C~David Levermore.
\newblock Fluid dynamic limits of kinetic equations ii convergence proofs for
  the boltzmann equation.
\newblock {\em Communications on pure and applied mathematics}, 46(5):667--753,
  1993.

\bibitem{quastel1998lattice}
Jeremy Quastel and H-T Yau.
\newblock Lattice gases, large deviations, and the incompressible navier-stokes
  equations.
\newblock {\em Annals of mathematics}, pages 51--108, 1998.

\bibitem{bandak2022dissipation}
Dmytro Bandak, Nigel Goldenfeld, Alexei~A Mailybaev, and Gregory Eyink.
\newblock Dissipation-range fluid turbulence and thermal noise.
\newblock {\em Physical Review E}, 105(6):065113, 2022.

\bibitem{bell_nonaka_garcia_eyink_2022}
John~B. Bell, Andrew Nonaka, Alejandro~L. Garcia, and Gregory Eyink.
\newblock Thermal fluctuations in the dissipation range of homogeneous
  isotropic turbulence.
\newblock {\em Journal of Fluid Mechanics}, 939:A12, 2022.

\bibitem{jensen2000large}
Leif~Harold Jensen.
\newblock {\em Large deviations of the asymmetric simple exclusion process in
  one dimension}.
\newblock PhD thesis, New York University, 2000.

\bibitem{varadhan2004large}
Srinivasa~RS Varadhan.
\newblock Large deviations for the asymmetric simple exclusion process.
\newblock In {\em Stochastic analysis on large scale interacting systems},
  volume~39 of {\em Adv. Stud. Pure Math.}, pages 1--28. Mathematical Society
  of Japan, Tokyo, 2004.

\bibitem{quastel2021hydrodynamic}
Jeremy Quastel and Li-Cheng Tsai.
\newblock Hydrodynamic large deviations of tasep.
\newblock {\em arXiv preprint arXiv:2104.04444}, 2021.

\bibitem{lejan2002integration}
Yves Le~Jan and Olivier Raimond.
\newblock Integration of {B}rownian vector fields.
\newblock {\em Ann. Probab.}, 30(2):826--873, 2002.

\bibitem{lejan2004flows}
Yves Le~Jan and Olivier Raimond.
\newblock Flows, coalescence and noise.
\newblock {\em Ann. Probab.}, 32(2):1247--1315, 2004.

\bibitem{eyink2015spontaneous}
Gregory~L Eyink and Theodore~D Drivas.
\newblock Spontaneous stochasticity and anomalous dissipation for {B}urgers
  equation.
\newblock {\em J. Stat. Phys.}, 158:386--432, 2015.

\bibitem{karr1975weak}
Alan~F Karr.
\newblock Weak convergence of a sequence of markov chains.
\newblock {\em Zeitschrift f{\"u}r Wahrscheinlichkeitstheorie und Verwandte
  Gebiete}, 33(1):41--48, 1975.

\bibitem{taylor1935statistical}
GI~Taylor.
\newblock Statistical theory of turbulence, i.
\newblock {\em Proceedings of the Royal Society of London, Series A,
  Mathematical and Physical Sciences}, 151(873):421--444, 1935.

\bibitem{suzuki2005energy}
Eijiro Suzuki, Tohru Nakano, Naoya Takahashi, and Toshiyuki Gotoh.
\newblock Energy transfer and intermittency in four-dimensional turbulence.
\newblock {\em Physics of Fluids}, 17(8), 2005.

\bibitem{gotoh2007statistical}
Toshiyuki Gotoh, Yusaku Watanabe, Yoshitaka Shiga, Tohru Nakano, and Eijiro
  Suzuki.
\newblock Statistical properties of four-dimensional turbulence.
\newblock {\em Physical Review E—Statistical, Nonlinear, and Soft Matter
  Physics}, 75(1):016310, 2007.

\bibitem{yamamoto2012local}
T~Yamamoto, H~Shimizu, T~Inoshita, T~Nakano, and Toshiyuki Gotoh.
\newblock Local flow structure of turbulence in three, four, and five
  dimensions.
\newblock {\em Physical Review E—Statistical, Nonlinear, and Soft Matter
  Physics}, 86(4):046320, 2012.

\bibitem{berera2020homogeneous}
Arjun Berera, Richard~DJG Ho, and Daniel Clark.
\newblock Homogeneous isotropic turbulence in four spatial dimensions.
\newblock {\em Physics of Fluids}, 32(8), 2020.

\bibitem{boffetta2012two}
Guido Boffetta and Robert~E Ecke.
\newblock Two-dimensional turbulence.
\newblock {\em Annual review of fluid mechanics}, 44(1):427--451, 2012.

\bibitem{kadanoff1989automata}
Leo~P Kadanoff, Guy~R McNamara, and Gianluigi Zanetti.
\newblock From automata to fluid flow: Comparisons of simulation and theory.
\newblock {\em Physical Review A}, 40(8):4527, 1989.

\bibitem{vansaarloos1982non}
W~Van~Saarloos, D~Bedeaux, and P~Mazur.
\newblock Non-linear hydrodynamic fluctuations around equilibrium.
\newblock {\em Physica A: Statistical Mechanics and its Applications},
  110(1-2):147--170, 1982.

\bibitem{zubarev1983statistical}
DN~Zubarev and VG~Morozov.
\newblock Statistical mechanics of nonlinear hydrodynamic fluctuations.
\newblock {\em Physica A: Statistical Mechanics and its Applications},
  120(3):411--467, 1983.

\bibitem{morozov1984langevin}
VG~Morozov.
\newblock On the langevin formalism for nonlinear and nonequilibrium
  hydrodynamic fluctuations.
\newblock {\em Physica A: Statistical Mechanics and its Applications},
  126(3):443--460, 1984.

\bibitem{espanol1998stochastic}
Pep Espa{\~n}ol.
\newblock Stochastic differential equations for non-linear hydrodynamics.
\newblock {\em Physica A: Statistical Mechanics and its Applications},
  248(1-2):77--96, 1998.

\bibitem{espanol2009microscopic}
Pep Espa{\~n}ol, Jes{\'u}s~G Anero, and Ignacio Z{\'u}{\~n}iga.
\newblock Microscopic derivation of discrete hydrodynamics.
\newblock {\em The Journal of chemical physics}, 131(24):244117, 2009.

\bibitem{eyink1990dissipation}
Gregory~L Eyink.
\newblock Dissipation and large thermodynamic fluctuations.
\newblock {\em Journal of statistical physics}, 61:533--572, 1990.

\bibitem{dezarate2006hydrodynamic}
J.O. de~Zarate and J.~Sengers.
\newblock {\em Hydrodynamic Fluctuations in Fluids and Fluid Mixtures}.
\newblock Elsevier Science, 2006.

\bibitem{mcmullen2022navier}
Ryan~M McMullen, Michael~C Krygier, John~R Torczynski, and Michael~A Gallis.
\newblock Navier-stokes equations do not describe the smallest scales of
  turbulence in gases.
\newblock {\em Physical Review Letters}, 128(11):114501, 2022.

\bibitem{sreenivasan1984scaling}
Katepalli~R Sreenivasan.
\newblock On the scaling of the turbulence energy dissipation rate.
\newblock {\em The Physics of fluids}, 27(5):1048--1051, 1984.

\bibitem{kaneda2003energy}
Yukio Kaneda, Takashi Ishihara, Mitsuo Yokokawa, Ken’ichi Itakura, and Atsuya
  Uno.
\newblock Energy dissipation rate and energy spectrum in high resolution direct
  numerical simulations of turbulence in a periodic box.
\newblock {\em Physics of Fluids}, 15(2):L21--L24, 2003.

\bibitem{bedrossian2019sufficient}
Jacob Bedrossian, Michele Coti~Zelati, Samuel Punshon-Smith, and Franziska
  Weber.
\newblock A sufficient condition for the kolmogorov 4/5 law for stationary
  martingale solutions to the 3d navier--stokes equations.
\newblock {\em Communications in Mathematical Physics}, 367:1045--1075, 2019.

\bibitem{iyer2023dependence}
Kartik~P Iyer.
\newblock Dependence of the asymptotic energy dissipation on third-order
  velocity scaling.
\newblock {\em Physical Review Fluids}, 8(8):L082601, 2023.

\bibitem{iyer2024zeroth}
Kartik Iyer, Theodore Drivas, Gregory Eyink, and Katepalli Sreenivasan.
\newblock Is the zeroth law valid for homogeneous and isotropic turbulence of
  incompressible fluids?
\newblock Proc. Nat. Acad. Sci.,submitted, 2024.

\bibitem{eyink1995besov}
Gregory~L Eyink.
\newblock Besov spaces and the multifractal hypothesis.
\newblock {\em Journal of statistical physics}, 78:353--375, 1995.

\bibitem{anselmet1984high}
F~Anselmet, Yl~Gagne, EJ~Hopfinger, and RA~Antonia.
\newblock High-order velocity structure functions in turbulent shear flows.
\newblock {\em Journal of Fluid Mechanics}, 140:63--89, 1984.

\bibitem{chen2005anomalous}
SY~Chen, B~Dhruva, S~Kurien, KR~Sreenivasan, and MA~Taylor.
\newblock Anomalous scaling of low-order structure functions of turbulent
  velocity.
\newblock {\em Journal of Fluid Mechanics}, 533:183--192, 2005.

\bibitem{iyer2020scaling}
Kartik~P Iyer, Katepalli~R Sreenivasan, and PK~Yeung.
\newblock Scaling exponents saturate in three-dimensional isotropic turbulence.
\newblock {\em Physical Review Fluids}, 5(5):054605, 2020.

\bibitem{buaria2020dissipation}
Dhawal Buaria and Katepalli~R Sreenivasan.
\newblock Dissipation range of the energy spectrum in high reynolds number
  turbulence.
\newblock {\em Physical Review Fluids}, 5(9):092601, 2020.

\bibitem{gorbunova2020analysis}
Anastasiia Gorbunova, Guillaume Balarac, Micka{\"e}l Bourgoin, L{\'e}onie
  Canet, Nicolas Mordant, and Vincent Rossetto.
\newblock Analysis of the dissipative range of the energy spectrum in grid
  turbulence and in direct numerical simulations.
\newblock {\em Physical Review Fluids}, 5(4):044604, 2020.

\bibitem{mcmullen2023thermal}
RM~McMullen, JR~Torczynski, and MA~Gallis.
\newblock Thermal-fluctuation effects on small-scale statistics in turbulent
  gas flow.
\newblock {\em Physics of Fluids}, 35(1):011705, 2023.

\bibitem{frisch1995turbulence}
Uriel Frisch.
\newblock {\em Turbulence: the legacy of AN Kolmogorov}.
\newblock Cambridge university press, 1995.

\bibitem{brezis2010functional}
H.~Brezis.
\newblock {\em Functional Analysis, Sobolev Spaces and Partial Differential
  Equations}.
\newblock Universitext. Springer New York, 2010.

\bibitem{boyer2012mathematical}
F.~Boyer and P.~Fabrie.
\newblock {\em Mathematical Tools for the Study of the Incompressible
  Navier-Stokes Equations and Related Models}.
\newblock Applied Mathematical Sciences. Springer New York, 2012.

\bibitem{schmeisser1987topics}
H.J. Schmeisser and H.~Triebel.
\newblock {\em Topics in Fourier Analysis and Function Spaces}, volume~42 of
  {\em Mathematik und ihre Anwendungen in Physik und Technik}.
\newblock Geest \& Portig, Leipzig, 1987.

\bibitem{sawano2018theory}
Y.~Sawano.
\newblock {\em Theory of Besov Spaces}, volume~56 of {\em Developments in
  Mathematics}.
\newblock Springer, Singapore, 2018.

\bibitem{veraar2008besov}
M~Veraar and T~Hytonen.
\newblock On {B}esov regularity of brownian motions in infinite dimensions.
\newblock {\em Prob. and Math. Stat.}, 28(1):143--162, 2008.

\bibitem{fernique1970integrabilite}
Xavier Fernique.
\newblock Int{\'e}grabilit{\'e} des vecteurs gaussiens.
\newblock {\em CR Acad. Sci. Paris S{\'e}r. AB}, 270:A1698--A1699, 1970.

\bibitem{baldi1992large}
Paolo Baldi, G~Ben Arous, and G{\'e}rard Kerkyacharian.
\newblock Large deviations and the {S}trassen theorem in {H}{\"o}lder norm.
\newblock {\em Stochastic processes and their applications}, 42(1):171--180,
  1992.

\bibitem{glimm2012quantum}
J.~Glimm and A.~Jaffe.
\newblock {\em Quantum Physics: A Functional Integral Point of View}.
\newblock Springer, New York, 2012.

\end{thebibliography}
\end{document}